\documentclass[
aps,
pra,
showpacs
amsmath,
amssymb,
onecolumn,
floatfix,
lengthcheck,
superscriptaddress
]{revtex4-1}

\usepackage{graphicx}
\usepackage{mathtools}
\usepackage{color}
\usepackage{braket}
\usepackage{hyperref}
\usepackage[normalem]{ulem}
\usepackage[english]{babel}
\usepackage{blindtext}
\usepackage{lipsum}  
\usepackage{amsthm}
\usepackage{tabularx}
\usepackage{bm}
\usepackage{dsfont}
\usepackage{bbold}

\begin{document}

\title{Ab-initio non-relativistic quantum electrodynamics: Bridging quantum chemistry and quantum optics from weak to strong coupling}

  \author{Christian Sch\"afer}
  \email[Electronic address:\;]{christian.schaefer@mpsd.mpg.de}
  \affiliation{Max Planck Institute for the Structure and Dynamics of Matter and Center for Free-Electron Laser Science \& Department of Physics, Luruper Chaussee 149, 22761 Hamburg, Germany}
  \author{Michael Ruggenthaler}
  \email[Electronic address:\;]{michael.ruggenthaler@mpsd.mpg.de}
  \affiliation{Max Planck Institute for the Structure and Dynamics of Matter and Center for Free-Electron Laser Science \& Department of Physics, Luruper Chaussee 149, 22761 Hamburg, Germany}
  \author{Angel Rubio}
  \email[Electronic address:\;]{angel.rubio@mpsd.mpg.de}
  \affiliation{Max Planck Institute for the Structure and Dynamics of Matter and Center for Free-Electron Laser Science \& Department of Physics, Luruper Chaussee 149, 22761 Hamburg, Germany}

\date{\today}

\begin{abstract}
\noindent
By applying the Born-Huang expansion, originally developed for coupled nucleus-electron systems, to the full nucleus-electron-photon Hamiltonian of non-relativistic quantum electrodynamics (QED) in the long-wavelength approximation, we deduce an exact set of coupled equations for electrons on photonic energy surfaces and the nuclei on the resulting polaritonic energy surfaces. This theory describes seamlessly many-body interactions between nuclei, electrons and photons including the quantum fluctuation of the electromagnetic field and provides a proper first-principle framework to describe QED-chemistry phenomena, namely polaritonic and cavity chemistry effects. Since the photonic surfaces and the corresponding non-adiabatic coupling elements can be solved analytically, the resulting expansion can be brought into a compact form which allows us to analyze aspects of coupled nucleus-electron-photon systems in a simple and intuitive manner.
Furthermore, we discuss structural differences between the exact quantum treatment and Floquet theory, show how existing implementations of Floquet theory can be adjusted to adhere to QED and highlight how standard drawbacks of Floquet theory can be overcome. We then highlight, by assuming that the relevant photonic frequencies of a prototypical cavity QED experiment are in the energy range of the electrons, how from this generalized Born-Huang expansion an adapted Born-Oppenheimer approximation for nuclei on polaritonic surfaces can be deduced. This form allows a direct application of first-principle methods of quantum chemistry such as coupled-cluster or configuration interaction approaches to QED chemistry. By restricting the basis set of this generalized Born-Oppenheimer approximation we furthermore bridge quantum chemistry and quantum optics by recovering simple models of coupled matter-photon systems employed in quantum optics and polaritonic chemistry. We finally highlight numerically that simple few-level models can lead to physically wrong predictions, even in weak-coupling regimes, and show how the presented derivations from first principles help to check and derive physically reliable simplified models.
\end{abstract}

\date{\today}

\maketitle

\section{Introduction}
\noindent
In the last decade tremendous experimental advances have allowed to investigate and control complex many-body systems strongly coupled to photons~\cite{ebbesen2016,sukharev2017,yoshihara2017superconducting}. In such situations of strong light-matter interactions novel physical effects can be observed such as symmetry protected collisions of strongly interacting photons~\cite{thompson2017}, Bose-Einstein condensation~\cite{kasprzak2006} and room-temperature polariton lasing~\cite{cohen2010} of exciton-polaritons, or even the control of the energy-levels in living bacteria~\cite{coles2017}. Such dramatic changes of physical and chemical properties of many-body systems can even be observed if no real photons are present and it is only the vacuum of, for example, an optical cavity that the matter couples to. Examples are a change of chemical reactivity under strong coupling to the vacuum electromagnetic field~\cite{thomas2016}, different transition states in gas phase and cavity~\cite{cs2018transition} and multiple-Rabi splittings under vibrational strong coupling~\cite{george2016}. Such experimental results highlight that disregarding the photonic degrees of freedom when calculating chemical and physical properties of many-body systems (as usually done in quantum chemistry and solid-state physics) can become inadequate when we change the bare electromagnetic vacuum, for example, by an optical cavity or nanoplasmonic devices, especially for strong light-matter coupling. In such cases we have to take into account electronic, nuclear and photonic degrees of freedom at the same time~\cite{ruggenthaler2018a}. Most interestingly, the intricate interplay between these basic constituents of matter can even prevail at room temperature and under ambient conditions, making such strong light-matter coupling situations especially interesting for quantum technologies. Besides possible applications in the context of room-temperature quantum-information technologies, the possibility to design strong-coupling based chemical reactors is intriguing. Ideally, a specific change in the electromagnetic vacuum would allow to change reactions without heating, which is a common drawback when lasers are used to control chemistry.\\

\noindent
While the resulting equations that one would need to solve in principle are well-known~\cite{spohn2004, ruggenthaler2014, ruggenthaler2018a}, they are numerically unfeasible in terms of full many-body wavefunctions. One way to make these equations numerically tractable is by extending methods of many-body theory such as current-density/density-functional theory~\cite{dreizler2012, ullrich2011} or Green's function methods~\cite{fetter2003, stefanucci2013} to coupled matter-photon systems~\cite{rajagopal1994, ruggenthaler2011b, tokatly2013, ruggenthaler2014, ruggenthaler2015, melo2015}. While first ab-initio results for coupled matter-photon problems are available~\cite{pellegrini2015, flick2017ab,flick2018light} so far most calculations have used simplified descriptions based on quantum-optical models like the Dicke or Tavis-Cummings model~\cite{tavis1968exact,garraway2011}. Although these models can be derived under certain assumptions from higher-lying theories~\cite{galego2015} and are able to reproduce well certain aspects of the experimental results~\cite{galego2016,du2017}, their limitations when applied outside of traditional quantum-optical situations are as of yet not well explored. Besides others, such models often ignore so-called dark states~\cite{xiang2017twodim}, do often not capture the diamagnetic shift~\cite{todorov2010, george2016} or lead to questionable predictions concerning a superradiant phase transition~\cite{viehmann2011superradiant}. Furthermore, most currently employed models~\cite{feist2017polaritonic, luk2017multiscale, ribeiro2018polariton} assume that the individual constituents of the physical ensemble that exhibits strong coupling remain unaffected. This is in contrast to experimental results that show that also the individual constituents can be affected~\cite{thomas2016, chikkaraddy2016, benz2016single, wang2017coherent, kazuma2018real}. To investigate these possible limitations as well as to provide a consistent way of how to improve shortcomings is especially timely, since such models have been used to predict interesting new effects~\cite{galego2017, martinez2017} in the context of the emerging field of polaritonic chemistry and they form the basis of our current understanding of strong light-matter interactions. Furthermore, scrutinizing these quantum-optical models from a quantum-chemical perspective could also help to analyze long-standing problems of quantum optics such as the realizability of a superradiant phase transition as predicted by the Dicke model~\cite{knight1978, vukics2012}.\\

\noindent
In this work we employ an unbiased and practical first-principles description based on non-relativistic quantum electrodynamics (QED)~\cite{spohn2004, ruggenthaler2014, ruggenthaler2018a}, highlight how this allows us to employ highly-accurate quantum-chemical methods to analyze and simulate general situations of light-matter interactions and scrutinize paradigmatic quantum-optical models. In this way we illustrate shortcomings of usual models, provide a consistent way to improve their reliability and identify connections between different situations of strong light-matter interactions, for example, due to a high-Q cavity or plasmonic nanostructures, in periodic systems or due to external driving. The presented framework therefore does not only allow to understand and predict effects of strong light-matter interaction from first principles, but can also serve as a guide to exchange ideas between different settings of light-matter interactions and even between different fields of modern quantum physics. We do so by performing the Born-Huang expansion~\cite{born1927,born1954dynamical} of the coupled nucleus-electron-photon wavefunction, which allows us to exactly rewrite the eigenvalue problem as a set of coupled equations for nuclear, electronic and photonic degrees of freedom. In contrast to previous approaches we do not combine the photonic degrees of freedom with the nuclei~\cite{flick2017b} or single out the photonic contribution~\cite{hoffmann2018light}, but group them with the electrons. While this gives rise to photonic potential-energy surface and non-adiabatic couplings for the electrons and thus also changes the usual electronic surfaces to polaritonic (electron-photon quasi-particles) surfaces for the nuclei, the photonic part can be solved analytically with the help of a shifted harmonic oscillator basis. This allows us to rewrite the Born-Huang expansion for coupled nucleus-electron-photon systems in a compact form. In this form the importance of the often ignored photon-mediated dipole self-energy term~\cite{rokaj2017light} (connected to the diamagnetic shift and superradiant phase transition) becomes evident and we show how photonic observables can be constructed from the Born-Huang expansion. This exact QED expansion shares certain similarities with the Floquet approach~\cite{shirley1965solution,gao2000nonperturbative}. We present how existing Floquet implementations can be adjusted to adhere to QED~\cite{ronca2018twodimensional} and thus avoid the drawbacks of standard Floquet theory. This also allows to import ideas from Floquet engineering~\cite{bukov2015universal} to polaritonic chemistry. Then, by assuming that the kinetic-energy contributions of the nuclei are small and that the frequency of the relevant photon modes is in the energy range of the electronic excitations, for example, due to an optical high-Q cavity, we deduce an adapted Born-Oppenheimer approximation (the explicit polariton approximation) which shares certain similarities with the approach presented in Ref.~\cite{galego2015} but guarantees the stability of the system. In this form the application of quantum-chemical methods able to tackle strong correlation in molecules or solids such as coupled-cluster or configuration-interaction approaches to coupled light-matter systems becomes straightforward. By further restricting the basis expansion for the electronic and the photonic subsystem we then arrive at a simple few-level approximation that resembles often employed model Hamiltonians. We then scrutinize the resulting few-level approximations for a numerically exactly solvable electron-photon problem and highlight how in the weak-coupling regime some integrated quantities like the energies are well reproduced but photon occupation and real-space quantities like the density can be qualitatively wrong. By including only a few more states the results improve considerably provided one increases the electronic and photonic basis sets consistently. In the strong- to ultra-strong-coupling regime, however, many more states need to be included for the integrated quantities to become accurate. This highlights that for strongly coupled problems the bare, that is, uncoupled, basis expansion is not very efficient and results based on a bare description, for example, using the standard Born-Oppenheimer states and dipole-coupling elements, become unreliable.\\

\noindent
This paper is structured as follows: In Sec.~\ref{theory} we introduce the basic Hamiltonian, present the Born-Huang expansion and show the analytic solution of the photonic subsystem. In Sec~\ref{implications} we then discuss the physical implications of the Born-Huang expansion and derive the explicit polariton approximation, which is a generalization of the quantum-chemical Born-Oppenheimer approximation. In Sec.~\ref{connection_qo} we derive the single photon polariton approximation with the help of a bare basis expansion and connect to quantum-optical models. Next, in Sec.~\ref{numchecks} we then give numerical details and present the accuracy of few-level approximations for the case of a GaAs quantum ring in a high-Q cavity. Finally, we conclude and give an outlook in Sec.~\ref{summary}. In the appendix we further discuss how the coupling strength effectively changes in collective systems and show how the resulting framework can also be used to consider periodic or time-dependent systems.

\section{Theory}\label{theory}
\noindent
We start by presenting the basic QED non-relativistic Hamiltonian for nuclei, electrons and photons that we consider. Then we perform the Born-Huang expansion in terms of nuclear, conditional electronic and photonic wavefunctions. Since the Hamiltonian is spin independent (we ignore the fine-structure influence by spin-orbit and spin-to-magnetic-field interactions) we can discard for notational simplicity and without loss of generality the spin degrees of freedom of the wavefunctions. By properly symmetrizing the wavefunctions at the end the physical eigenstates can be found. We further use atomic units throughout the work.

\subsection{Pauli-Fierz Hamiltonian in the long-wavelength limit}
\noindent
Let us assume that for a coupled matter-photon system the relevant photon modes have wavelengths that are large compared to the typical size of a matter subsystem, for example, a molecule. This typically happens, for instance, if we are interested in ground or excited states and not consider long-time dynamics and scattering states that can spread over space. By definition these states are exponentially localized~\cite{spohn2004} and in this case an approximation of the full Pauli-Fierz Hamiltonian of non-relativistic QED in Coulomb gauge~\cite{spohn2004,ruggenthaler2014,ruggenthaler2018a}, where we neglect the spatial dependence of the photon modes in the length form~\cite{pellegrini2015,flick2015,flick2017ab}
\begin{align}\label{hamiltonian}
\hat{H} &= \hat{H}_{n} + \hat{H}_{e} + \hat{H}_{ne} + \hat{H}_{p} + \hat{H}_{ep} + \hat{H}_{np}
\end{align}
is known to be accurate.
Here the nuclear Hamiltonian for $N_n$ nuclei
\begin{align*}
\hat{H}_{n} &= \hat{T}_n + \hat{W}_{nn} = \sum_{j=1}^{N_n} -\frac{1}{2 M_j}\nabla_{\textbf{R}_j}^2 + \frac{1}{2}\sum_{i,j\neq i}^{N_n} \frac{Z_{i}Z_{j}}{\vert \hat{\textbf{R}}_i - \hat{\textbf{R}}_j \vert}
\end{align*}
consists of the sum over all kinetic components for each nucleus $j$ with effective nuclear mass $M_j$ and Coulombic nucleus-nucleus interaction $\hat{W}_{nn}$ with $Z_j$ being the effective positive nuclear charges. The electronic Hamiltonian for $N_e$ electrons
\begin{align*}
\hat{H}_{e} &= \hat{T}_e +\hat{W}_{ee} = -\frac{1}{2m_e}\sum_{j=1}^{N_e} \nabla_{\textbf{r}_j}^2 + \frac{1}{2}\sum_{i,j\neq i}^{N_e} \frac{1}{\vert \hat{\textbf{r}}_i - \hat{\textbf{r}}_j \vert}
\end{align*}
includes similarly the corresponding sum over electronic kinetic components with the electron mass $m_e$ and the Coulomb electron-electron interaction. The nuclear-electron interaction is given accordingly by
\begin{align*}
\hat{H}_{ne} =  -\sum_{j=1}^{N_n}\sum_{i=1}^{N_e}\frac{Z_{j}}{\vert \hat{\textbf{r}}_i - \hat{\textbf{R}}_j \vert} ~.
\end{align*}
Further, the photonic contribution for $M_p$ modes is then given by
\begin{align*}
\hat{H}_{p} + \hat{H}_{ep} + \hat{H}_{np} = \frac{1}{2}\sum\limits_{\alpha=1}^{M_p}\left[ \hat{p}_{\alpha}^2 + \omega_{\alpha}^2 \left(\hat{q}_{\alpha} -\frac{\boldsymbol\lambda_{\alpha}}{\omega_{\alpha}} \cdot \hat{\textbf{R}} \right)^2 \right], 
\end{align*}
which incorporates the total dipole $\hat{\textbf{R}} = \sum_{j=1}^{N_{e}}\hat{\textbf{r}}_{j}-\sum_{j=1}^{N_{n}}Z_j\hat{\textbf{R}}_{j}$
of electrons and nuclei~\footnote{We restrict our self to the long-wavelength approximation as it is a standard assumption in cavity QED. Nevertheless, the derivations presented here are conceptually not restricted to this simplified choice. We will briefly note qualitative changes where relevant.}. Here $M_p$ is a finite but arbitrarily large amount of photon modes which are the most relevant modes (see also Subsec.~\ref{nboa}) but in principle run from the fundamental mode of our arbitrarily large but for simplicity finite quantization volume~\footnote{We could straightaway also use the full infinite space~\cite{spohn2004}, but this will just make the notation unnecessarily complicated, since we will reduce the photon modes in the course of the paper to a few effective ones.} up to a maximum sensible frequency, for example, an ultra-violet cut-off at rest mass energy of the electrons. The quantized oscillators representing the photonic system consists of the canonical coordinate corresponding to the displacement field $\hat{q}_\alpha = \frac{1}{\sqrt{2\omega_\alpha}}(\hat{a}^\dagger_\alpha + \hat{a}_\alpha) $ and its conjugate momentum $ \hat{p}_\alpha = -i \sqrt{\frac{\omega_\alpha}{2}} (\hat{a}_\alpha-\hat{a}^\dagger_\alpha)  \equiv -i \tfrac{\partial}{\partial \hat{q}_{\alpha}}$ as presented in \cite{pellegrini2015,flick2017ab} with $\left[\hat{q}_\alpha,\hat{p}_{\alpha'}\right] = i\delta_{\alpha\alpha'}$. The fundamental coupling strength $\boldsymbol\lambda_\alpha = \lambda_{\alpha} \textbf{e}_\alpha$ describes the coupling between the total dipole and the photonic mode $\alpha$ with wavevector $\textbf{k}_\alpha$ and transversal polarization vector $ \textbf{e}_\alpha$. Here the coupling strength 
\begin{align}\label{fundamentalcoupling}
\boldsymbol\lambda_\alpha = \sqrt{4\pi} S_\alpha(\textbf{r}) \textbf{e}_\alpha
\end{align}
depends on the form of the mode functions $S_\alpha(\textbf{r})$ and the chosen reference point for our matter subsystem~\cite{ruggenthaler2014,pellegrini2015,flick2017ab}. If we consider free space (as usually done in quantum chemistry) then they will be the usual exponentials, while if we consider a system in an, e.g., optical cavity, their form might be very different. This different form can then lead to an enhanced coupling of a specific mode with respect to the usual free-space case. This increase of the fundamental coupling is an inherent feature of the physical set-up, for example, the form and nature of the cavity, and cannot directly be amplified by the number of charged particles. We discuss this in some more detail in App.~\ref{colcoupling} and highlight how the number of emitters can, however, enhance the effective coupling strength when a reduced description is employed. In the case of a free-space problem the photon modes that give rise to the radiative losses, i.e., they constitute the photon bath of the matter subsystem, are then usually only taken into account by renormalizing the bare masses $m_e$ and $M_j$ of the charged particles. Thus instead of the photonic degrees of freedom, one uses the ``physical'' masses of the particles that contain the bare and the electromagnetic masses. This procedure is highly accurate when time-independent problems (the focus of this work) are concerned, and only on longer time scales the missing dissipation due to the photons is known to become relevant. We note at this point that all the dissipation due to phonons is still included exactly since we treat the nuclei explicitly.\\
\noindent
In the case that we now change the modes from free space to those that are associated with, e.g., a cavity, we can in principle perform the same procedure. We could keep many modes such that we describe the openness of the cavity, as has been demonstrated in practice recently in Ref.~\cite{flick2018light}, or we reduce to slightly changed ``physical'' masses due to the changed photon field. In the following we do not intend to simulate the continuum of states by explicitly treating many modes, since we will be mainly interested in static properties. Therefore our working assumption will be that we keep the well-established physical masses of the particles, e.g.,  $m_e=1$, and treat the effect of the changes in the photon field due to, e.g., a nanoplasmonic device, by keeping a few of the enhanced modes. This allows us to keep the effect of the openness of the photonic environment as a mass-renormalization. The numerical investigations in Sec.~\ref{numchecks} then refer to such a situation, where on top we further assume quite strong effective couplings (for instance due to ensemble effects, as discussed in App.~\ref{colcoupling}), approachable for example in circuit QED~\cite{niemczyk2010circuit,yoshihara2017superconducting}. Wherever we refer to weak or strong coupling, the notion is motivated by the effect of the light-matter coupling on the matter subsystem and the strength of the Rabi-splitting. In the weak coupling regime the Rabi splitting is small, and so are the changes in the matter subsystem with respect to the free-space case. But we point out that these changes are present even for explicit dissipation~\cite{spohn2004,flick2017,liberato2017virtual}, and we therefore do not make the standard assumption as done in, e.g., investigations of the amplified spontaneous emission rates due to the Purcell effect~\cite{purcell1995spontaneous}, that the matter subsystem remains unaffected. In the strong and ultra-strong coupling regime, where the Rabi splitting becomes very large, these changes become more pronounced.\\

\noindent
The self-polarization part in the photonic Hamiltonian $\frac{1}{2}(\boldsymbol\lambda_\alpha\cdot\hat{\textbf{R}})^2$ naturally arises in the length form to make the Hamiltonian bounded from below, which is a prerequisite to allow for ground states of an interacting light-matter many-body system~\cite{rokaj2017light}. In the Subsec.~\ref{relselfpol} we give a very intuitive physical picture for this abstract statement.
The self-polarization further renders the Hamiltonian invariant under translations provided we have a charge neutral system, that is, $\sum_{j=1}^{N_n}{Z_j} =N_e$. In the case that the total system is not charge neutral, the center-of-charge couples to the photonic field and for eigenstates a translation by $\bar{\textbf{R}}$, namely $\hat{\textbf{R}}  \rightarrow \hat{\textbf{R}}+\bar{\textbf{R}}$, leads to a trivial elongation of the photonic displacement by $\hat{q}_\alpha \rightarrow \hat{q}_\alpha-\frac{\boldsymbol\lambda_\alpha}{\omega_\alpha} \cdot \bar{\textbf{R}}$ (see also Sec.~\ref{photonsub}).

\noindent
For the above nucleus-electron-photon Hamiltonian we then want to determine the eigenfunctions~\footnote{To be precise, in the infinite space the full Hamiltonian has no eigenstates, just scattering states (generalized eigenstates). So one either has to enclose everything in a big but finite box or position the molecular system by some external potential, interpret the eigenstate equation in a weak sense or separate off the translationally and rotationally invariant parts as usually done in molecular physics. We will, as usually done, disregard this subtlety in the following for simplicity.}
\begin{align}\label{initialeigeneq}
\hat{H} \Psi_{i}(\textbf{R}_n,\textbf{r},  \textbf{q}) = E_i \Psi_{i}( \textbf{R}_n, \textbf{r}, \textbf{q})~.
\end{align}
with the many-body energies $E_i$, where $\textbf{R}_n, \textbf{r}$ and $\textbf{q}$ are collective variables defined as $\textbf{R}_n =  (\textbf{R}_1,\textbf{R}_2,...,\textbf{R}_{N_n})$, $\textbf{r} = ( \textbf{r}_1, \textbf{r}_2,..., \textbf{r}_{N_e})$ and $\textbf{q} = (q_{1},q_2,...,q_{M_p})$. Here we note that if we have all the eigenfunctions of the Hamiltonian we also have direct access to all temperature effects, since we can directly determine the canonical ensemble $\hat{\rho}=\exp(\hat{H}/k_{B}T) = \sum_{i=1}^{\infty} \exp(E_i/k_{B}T) \ket{\Psi_i}\bra{\Psi_i}$. Next we perform a Born-Huang expansion where we expand into subsystem wavefunctions.
This expansion can be performed in multiple different ways resulting in alternative physical interpretations and consequences for approximations. We will elaborate on their relevance and implications a little later. Here we use $ \chi^{\mu}_{i}(\textbf{R}_n)$ that represent nuclear wavefunctions and $\tilde{\Psi}_{\mu}(\{\textbf{R}_n\}, \textbf{r},  \textbf{q})$ the polaritonic components describing the correlated electron-photon system. We choose $\tilde{\Psi}_{\mu}$ such that they form an orthonormal basis in the electron-photon subsystem and assume that they depend in an yet unspecified parametric way on the positions of the nuclei $\{ \textbf{R}_n \}$. Due to this parametric dependence we will later find equations that couple the different subsystem wavefunctions. In a second step, we further expand the polaritonic wavefunction into electronic $\psi^k_{\nu}(\textbf{r}, \{\textbf{R}_n\})$ and photonic $\Phi_{k}(\textbf{q},\{\textbf{R}\})$ subsystem wavefunctions 
\begin{align}
\Psi_{i}(\textbf{R}_n, \textbf{r}, \textbf{q}) &=  \sum_{\mu = 0}^{\infty} \chi^{\mu}_{i}(\textbf{R}_n) \tilde{\Psi}_{\mu}(\{\textbf{R}_n\}, \textbf{r},  \textbf{q}) \label{fac1}\\
&= \sum_{\mu, k = 0}^{\infty} \chi^{\mu}_{i}(\textbf{R}_n) \psi^k_{\mu}(\textbf{r}, \{\textbf{R}_n\}) \Phi_{k}(\textbf{q},\{\textbf{R}\})\notag
\end{align}
We use here that in the photonic subspace we employ an orthonormal basis of wavefunctions $\Phi_{k}(\textbf{q},\{\textbf{R}\})$ that parametrically depend on the total dipole of the matter (nuclei and electrons) subsystem, that is, $\langle \Phi_{k'} \vert \Phi_{k} \rangle_p = \int d\textbf{q}\, \Phi_{k'}^*(\textbf{q},\{\textbf{R}\}) \Phi_{k}(\textbf{q},\{\textbf{R}\}) = \delta_{kk'}$, as well as electronic wavefunctions parametrically dependent on the position of the nuclei such that $\sum_{k=0}^{\infty}\langle  \psi^{k}_{\mu'} \vert \psi^{k}_{\mu}\rangle_e = \sum_{k=0}^{\infty}\int d\textbf{r}\psi^{k*}_{\mu'}(\textbf{r},\{\textbf{R}_n\}) \psi^{k}_{\mu}(\textbf{r},\{\textbf{R}_n\}) = \delta_{\mu \mu'}$. This allows to re-express the normalization as
\begin{align}
\label{normalizationfull}
\langle \Psi_{i} \vert \Psi_{i}\rangle &= \sum_{\mu', k' = 0}^{\infty} \sum_{\mu, k = 0}^{\infty} \langle \chi^{\mu'}_{i} \vert \chi^{\mu}_{i} \rangle_n \langle  \psi^{k'}_{\mu'} \vert \psi^{k}_{\mu} \rangle_e \langle \Phi_{k'} \vert \Phi_{k} \rangle_p \notag\\
&=\sum_{\mu,\mu' =0}^{\infty} \langle \chi^{\mu'}_{i} \vert \chi^{\mu}_{i} \rangle_n \sum_{k = 0}^{\infty}\langle  \psi^{k}_{\mu'} \vert \psi^{k}_{\mu}\rangle_e\notag\\
&=\sum_{\mu=0}^{\infty} \langle \chi^{\mu}_{i} \vert \chi^{\mu}_{i} \rangle_n = 1~.
\end{align}
Note that by the introduced expansion we find for each polaritonic eigenstate $\tilde{\Psi}_{\mu}$ electronic states $\psi^{k}_{\mu}$ associated to a photonic excitation $\Phi_{k}$. This shows that the electronic space is repeated with associated photonic excitations. We can see here a similarity to Floquet-theory~\cite{shirley1965solution,gao2000nonperturbative,kuchment2012floquet} (see Sec.~\ref{nboa} and \ref{floquet} for details) where a matter system driven by a classical external field is considered. In this case, by assuming periodic driving, a time-dependent problem can be rewritten as an eigenvalue problem in a Hilbert space including time. The resulting eigenvalue equation is unbounded from below which expresses itself by the Floquet block index $l \in \mathbb{Z}$ that is usually interpreted (despite the absence of actual photons in the mathematical formulation of the problem) as the number of photons involved in the process. The Floquet approach therefore has no well-defined ground state and allows for negative ``photon excitations'' which are often interpreted as the emission of photons. To identify the physically correct occupation of a ``photon-dressed'' system in Floquet theory is a very subtle issue of intense discussion in the community~\footnote{A well defined approach to collect occupations results from a time-dependent calculation and subsequent projection. In this case, Floquet theory reduces to a post-processing tool to analyze the complex driven many-body dynamics. Within this paper, we will discuss how a Floquet-type algorithm deduced from QED can be used in this context to obtain results without ambiguity.}. In contrast to Floquet-theory, for the fully coupled matter-photon problem the number of possible photonic excitations $k$ are bounded from below by $k=0$, which is the vacuum of the electromagnetic field.

\subsection{Coupled equations in separated Hilbert spaces}
\noindent
Let us next derive, by applying the full Hamiltonian \eqref{hamiltonian} on the discussed expansion, coupled equations for the parametrically dependent subsystem wavefunctions. 
We employ here that in configuration space $(\textbf{R}_n,\textbf{r}, \textbf{q})$ multiplication operators like potentials and interactions commute with the wavefunctions. This allows us to formally treat the dependence on other subsystems parametrically, for example, for the electronic wavefunction the electron-nucleus interaction becomes
\begin{align*}
&\sum_{j=1}^{N_n}\sum_{h=1}^{N_e}\frac{-Z_{j}}{\vert \hat{\textbf{r}}_h - \hat{\textbf{R}}_j \vert} \chi_i^\mu(\textbf{R}_n) \psi_\mu^k(\textbf{r},\{\textbf{R}_n\})\\
&= \chi_i^\mu(\textbf{R}_n)\sum_{j=1}^{N_n}\sum_{h=1}^{N_e}\frac{-Z_{j}}{\vert \hat{\textbf{r}}_h - \textbf{R}_j \vert} \psi_\mu^k(\textbf{r},\{\textbf{R}_n\})~.
\end{align*}
In a more physical interpretation, as long as the particles have no quantum-character, the coupling between different systems is purely recovered by their mean-values. This simple factorization is no longer valid if the Hamiltonian includes derivatives and therefore assigns a quantum-character to the particles. The kinetic contributions act on all functions, that is, there are non-vanishing contributions such as $ \left[\nabla_{\textbf{R}_j} \chi^{\mu}_{i}\right] \left[\nabla_{\textbf{R}_j} \psi^{k}_{\mu}\right] $, which explicitly couple nuclear, electronic and photonic degrees of freedom. The initial eigenvalue equation then becomes 
\begin{widetext}
\begin{align}
\begin{split}
\label{fullequ}
E_{i}\Psi_{i} &= \hat{H} \sum_{\mu, k = 0}^{\infty} \chi^{\mu}_{i}(\textbf{R}_n) \psi^k_{\mu}(\textbf{r}, \{\textbf{R}_n\}) \Phi_{k}(\textbf{q},\{\textbf{R}\})\\
&= \sum_{\mu, k = 0}^{\infty}\hat{H}_n \left[ \chi^{\mu}_{i}(\textbf{R}_n) \psi^k_{\mu}(\textbf{r}, \{\textbf{R}_n\}) \Phi_{k}(\textbf{q},\{\textbf{R}\})\right]
+ \sum_{\mu, k = 0}^{\infty} \chi^{\mu}_{i}(\textbf{R}_n) \left[\hat{H}_{e} + \hat{H}_{ne}(\{\textbf{R}_n\}) \right] \psi^k_{\mu}(\textbf{r},\{\textbf{R}_n\}) \Phi_{k}(\textbf{q},\{\textbf{R}\})\\
&+ \sum_{\mu, k = 0}^{\infty} \chi^{\mu}_{i}(\textbf{R}_n\boldsymbol) \psi^k_{\mu}(\textbf{r},\{\textbf{R}_n\})    \left[\hat{H}_{p} + \hat{H}_{ep}(\{\textbf{r}\}) + \hat{H}_{np}(\{\textbf{R}_n\})\right] \Phi_{k}(\textbf{q},\{\textbf{R}\})~,
\end{split}
\end{align}
\end{widetext}
where we suppressed the spatial dependences on the full wavefunction on the left hand side for notational convenience. This equation can be exactly decomposed after multiplication with $\psi^{l*}_{\nu} \Phi^*_{l}$ and a subsequent summation and integration, that is,
\begin{align*}
 \sum_{l=0}^{\infty} \int d\textbf{r} \psi^{l*}_{\nu}(\textbf{r}, \{\textbf{R}_n\}) \int d \textbf{q} \Phi^*_{l}(\textbf{q}, \{\textbf{R}\})~,
\end{align*}
into photonic, electronic and nuclear subspaces.
With the normalization defined as in Eq.~\eqref{normalizationfull}, we arrive at a reduced equation for the nuclear subspace
\begin{widetext}
\begin{align}
\label{fullseparated1}  E_i \chi^{\nu}_{i}(\textbf{R}_n) &=\hat{H}_{n}\chi^{\nu}_{i}(\textbf{R}_n)
\\
\label{fullseparated1p5} &- \frac{1}{2} \sum_{\mu,l,k=0}^{\infty} \sum_{j=1}^{N_n} \frac{-Z_j}{M_j} \left[ 2\langle \psi^l_{\nu} \vert \psi^k_{\mu} \rangle_e \nabla^{lk}\cdot \nabla_{\textbf{R}_j} + 2 \langle \psi^l_{\nu} \vert \nabla_{\textbf{R}_j}\vert \psi^k_{\mu} \rangle_e \cdot \nabla^{lk}
- \langle \psi^l_{\nu} \vert \psi^k_{\mu} \rangle_e \Delta^{lk} Z_j \right] \chi^{\mu}_{i}(\textbf{R}_n)  
\\
\label{fullseparated2} &- \frac{1}{2} \sum_{\mu,l=0}^{\infty} \sum_{j=1}^{N_n} \frac{1}{M_j} \left[ 2 \langle \psi^l_{\nu} \vert \nabla_{\textbf{R}_j}  \vert \psi^l_{\mu} \rangle_e \cdot \nabla_{\textbf{R}_j} + \langle \psi^l_{\nu} \vert \Delta_{\textbf{R}_j}  \vert \psi^l_{\mu} \rangle_e \right] \chi^{\mu}_{i}(\textbf{R}_n) 
\\
\label{fullseparated3} &+ \sum_{\mu, l, k= 0}^{\infty} \chi^{\mu}_{i}(\textbf{R}_n) \langle\psi^l_{\nu} \vert \left(
\left(\hat{H}_e + \hat{H}_{ne}(\{\textbf{R}_n\})\right) \delta_{lk} - \frac{1}{2} \left[ 2 \nabla^{lk} \cdot \sum_{j=1}^{N_e} \nabla_{\textbf{r}_j} + N_e \Delta^{lk} \right] \right)\vert \psi^k_{\mu}\rangle_e
\\
\label{fullseparated4} &+ \sum_{\mu,l, k = 0}^{\infty} \chi^{\mu}_{i}(\textbf{R}_n) \langle \psi^l_{\nu} \vert  \langle \Phi_{l} \vert\left[\hat{H}_{p} + \hat{H}_{ep}(\{\textbf{r}\}) + \hat{H}_{np}(\{\textbf{R}_n\})\right] \vert \Phi_{k} \rangle_{p} \vert\psi^k_{\mu} \rangle_{e},
\end{align}
where the photonic coupling elements are given by (see App.~\ref{analytic_coupling} for a detailed derivation)
\begin{align}
\nabla^{lk} &= \int d\textbf{q} \Phi_{l}^*(\textbf{q},\{\textbf{R}\}) \nabla_{R} \Phi_{k}(\textbf{q},\{\textbf{R}\}) \label{couplings2}
= -\sum\limits_{\alpha}^{M_p} \frac{\boldsymbol\lambda_\alpha}{\sqrt{2\omega_\alpha}} \left[ \sqrt{k_\alpha+1} \delta^\alpha_{l,k+1} - \sqrt{k_\alpha} \delta^\alpha_{l,k-1} \right] 
= \sum\limits_{\alpha}^{M_p} \nabla^{lk,\alpha} = -\nabla^{kl},
\end{align}
and
\begin{align}
\Delta^{lk} &= \int d\textbf{q} \Phi_{l}^*(\textbf{q},\{\textbf{R}\}) \Delta_{R} \Phi_{k}(\textbf{q},\{\textbf{R}\}) \label{couplings3}\\ &= + \sum\limits_{\alpha}^{M_p} \left(\frac{\boldsymbol\lambda_\alpha}{\sqrt{2\omega_\alpha}}\right)^2 \bigg[ -(2k_\alpha+1) \delta^\alpha_{l,k} + \sqrt{(k_\alpha+1)(k_\alpha+2)} \delta^\alpha_{l,k+2} + \sqrt{k_\alpha(k_\alpha-1)} \delta^\alpha_{l,k-2} \bigg]
+ \sum\limits_{\alpha',\alpha\neq \alpha'}^{M_p} \nabla^{lk,\alpha'} \cdot \nabla^{lk,\alpha} \notag\\
&= \sum\limits_{\alpha}^{M_p} \Delta^{lk,\alpha} + \sum\limits_{\alpha',\alpha\neq \alpha'}^{M_p} \nabla^{lk,\alpha'} \cdot \nabla^{lk,\alpha} = \Delta^{kl} ~.\notag
\end{align}
\end{widetext}
In this form we can identify three equations which have to be solved in a self-consistent manner in order to satisfy the above combined nucleus-electron-photon problem. The first one is given by the photonic equation of \eqref{fullseparated4}. Since the total dipole $\textbf{R}$ that shows up in the coupling to the photon subsystem wavefunction is given by merely the electronic and nuclear coordinates, for eigenstates with 
\begin{align}\label{phsub}
\begin{split}
&\left[\hat{H}_{p} + \hat{H}_{ep}(\{\textbf{r}\})+ \hat{H}_{np}(\{\textbf{R}_n\})\right] \Phi_{k}(\textbf{q},\{\textbf{R}\})\\
&= \varepsilon_k(\{\textbf{R}\}) \Phi_{k}(\textbf{q},\{\textbf{R}\})~,
\end{split}
\end{align}
the term \eqref{fullseparated4} simplifies to 
\begin{align}\label{fullseparated5}
\sum_{\mu,l, k = 0}^{\infty} \chi^{\mu}_{i}(\textbf{R}_n) \langle\psi^l_{\nu}(\{\textbf{R}_n\})\vert \varepsilon_k(\{\textbf{R}\}) \vert \psi^k_{\mu}(\{\textbf{R}_n\})\rangle_e   ~\delta_{lk}~.
\end{align}
The dependence of the photonic eigenvalue on the total dipole therefore leads to a photonic potential-energy surface $\varepsilon_k(\{\textbf{R}\}) \equiv \varepsilon_k(\{ \textbf{R}_n, \textbf{r}\})$. We denote the parametrically dependent photonic Hamiltonian as the \textit{photonic Born-Oppenheimer} Hamiltonian
\begin{align}
 \hat{H}_{BO}^{ph}(\{ \textbf{R}_n, \textbf{r}\}) = \frac{1}{2}\sum\limits_{\alpha}^{M_p}\left[ \hat{p}_{\alpha}^2 + \omega_{\alpha}^2 \left(\hat{q}_{\alpha} -\frac{\boldsymbol\lambda_{\alpha}}{\omega_{\alpha}} \cdot \textbf{R} \right)^2\right]. \label{bohamph}
\end{align}
We can then shift the photonic potential-energy surface into \eqref{fullseparated3} and define a \textit{photon-adapted electronic Born-Oppenheimer} Hamiltonian according to 
\begin{align}
\hat{H}_{BO}^l(\textbf{r},\{\textbf{R}_n\}) = \hat{H}_{e} + \hat{H}_{ne}(\textbf{r},\{\textbf{R}_n\}) + \varepsilon_{l}(\textbf{r},\{\textbf{R}_n\}) \label{boham1}~.
\end{align}
With this definition we then solve for the electronic eigenfunctions of \eqref{fullseparated3} including the photonic potential-energy surfaces as
\begin{align}\label{elsub}
&\hat{H}_{BO}^l(\textbf{r},\{\textbf{R}_n\}) \psi^l_{\mu}(\textbf{r}, \{\textbf{R}_n\}) \notag\\
&- \frac{1}{2} \sum_{k=0}^{\infty} \left[ 2 \nabla^{lk} \cdot \sum_{j=1}^{N_e} \nabla_{\textbf{r}_j} + N_e \Delta^{lk} \right] \psi^k_{\mu}(\textbf{r}, \{\textbf{R}_n\}) \notag\\
&= E_\mu(\{\textbf{R}_n\}) \psi^l_{\mu}(\textbf{r}, \{\textbf{R}_n\})~.
\end{align}
The remaining nuclear equation is now the combination of \eqref{fullseparated1}, \eqref{fullseparated1p5}, \eqref{fullseparated2} and the additional potential-energy surface $E_\mu(\{\textbf{R}_n\})$ from \eqref{elsub} such that
\begin{align}\label{nsub}
&\left[ \hat{H}_{n} + E_\nu(\textbf{R}_n) \right] \chi^{\nu}_{i}(\textbf{R}_n) \notag\\
&- \frac{1}{2} \sum_{\mu,l,k=0}^{\infty} \sum_{j=1}^{N_n} \frac{1}{M_j} \bigg(\bigg[ 2 (-Z_j) \langle \psi^l_{\nu} \vert \psi^k_{\mu} \rangle_e \nabla^{lk} \cdot  \nabla_{\textbf{R}_j} \notag
\\
& \quad+ 2 (-Z_j)\langle \psi^l_{\nu} \vert \nabla_{\textbf{R}_j}\vert \psi^k_{\mu} \rangle_e \cdot \nabla^{lk} + \langle \psi^l_{\nu} \vert \psi^k_{\mu} \rangle_e \Delta^{lk} Z_j^2 \bigg] \notag 
\\
&+ \delta_{kl}\left[ 2 \langle \psi^l_{\nu} \vert \nabla_{\textbf{R}_j}  \vert \psi^l_{\mu} \rangle_e \cdot \nabla_{\textbf{R}_j} + \langle \psi^l_{\nu} \vert \Delta_{\textbf{R}_j}  \vert \psi^l_{\mu} \rangle_e \right]\bigg) \chi^{\mu}_{i}(\textbf{R}_n) \notag
\\
&= E_i \chi^{\nu}_{i}(\textbf{R}_n)~.
\end{align}
We finally end up with three equations \eqref{phsub},\eqref{elsub} and \eqref{nsub} that have to be solved self-consistently. Their physical interpretation is that electrons move adiabatically on photonic energy-surfaces $ \varepsilon_{l}(\textbf{r},\{\textbf{R}_n\}) $ while excitations of the electronic system are coupled by photonic excitations $l$. The bilinear coupling \eqref{couplings2} transfers electronic momentum between different electronic states, mediated by photonic excitations. However,  within the long-wavelength approximation, the photon itself does not transfer momentum to the electron, see also App.~\ref{periodic}. The quadratic coupling \eqref{couplings3} in contrast is an energetic shift between eigenstates. The equivalence between the presented Born-Huang expansion and the Power-Zienau-Woolley transformation \cite{power1965introductory, andrews2018perspective} is elaborated in section \ref{pzwequiv}. In combination, \eqref{phsub} and \eqref{elsub} constitute the polaritonic subsystem which is interacting with the nuclei.
The nuclei move adiabatically on the polaritonic surfaces $E_\nu(\textbf{R}_n)$ with additional couplings mediated by photonic and electronic excitations as well as mixed electron-photon excitations. Bare non-adiabatic coupling elements without photons have to be adjusted, as for example discussed in Sec.~\ref{roothaanhall}, to account for novel contributions and the change of the electronic structure under photonic influence $\langle \psi^l_{\nu} \vert \nabla_{\textbf{R}_j}  \vert \psi^l_{\mu} \rangle$. The non-adiabatic couplings are often negligible as long as the Born-Oppenheimer surfaces $E_\nu(\textbf{R}_n)$ (in our case these are the polaritonic potential-energy surfaces) are energetically well separated but become relevant close to conical interactions \cite{kouppel1984multimode,kowalewski2015catching,galego2015,flick2017}. 
For the electron-photon subspace two important limiting cases arise. In the off-resonant case, the energy-replica of \eqref{phsub} will merely resemble a series of harmonic states on top of electronic surfaces without significant effect on the excited electronic state. Although the coupling might be sufficient to slightly distort quantities such as the electronic density, the excited state will not mix strongly with the photonic replica. In resonance, where the photon frequency is almost identical to an electronic transition, the non-adiabatic couplings become dominant and we find avoided crossings similar to the well-known electron-nuclear case. For the excited-state structure in the weak-coupling regime we therefore see explicitly why a single photon-mode close to resonance is of particular relevance while others have only a minor effect. Similarly, for the ground state in weak coupling the lowest mode is the most relevant one, as can be seen in the exact-exchange approximation to the electron-photon coupling that scales as $1/\omega$~\cite{pellegrini2015}. This can, however, change as we approach the ultra-strong coupling regime. Finally, we point out that a similar construction of coupled equations can be deduced also for the time-dependent case (see App.~\ref{timedependent}). Here, however, we can only expect our considerations to be accurate for a limited amount of propagation time. After a certain time the fact that we treat the continuum of photon modes only approximately will become apparent and beyond this point more advanced open-quantum-system approaches become necessary.

\subsection{Analytic solution of the photonic subspace}\label{photonsub}
\noindent
So far we have reformulated the high-dimensional problem of a correlated nucleus-electron-photon system into a set of three lower-dimensional, but coupled equations. The solution of these coupled equations is of course still as hard (maybe even harder) than the original problem. In order to make the problem numerically tractable we either need to introduce approximations or we provide analytical solutions to parts of these equations. In this subsection we will do the latter and bring the problem into a more compact form. We use that in the photonic equations \eqref{phsub} and \eqref{phsubtd}, respectively, the parametric dependence on the total dipole allows for an analytic solution. The dipole merely introduces a coherent shift in the photonic harmonic oscillator equations. Let us elaborate on this briefly in a generalized time-dependent picture as it will allow us to extend the Born-Huang framework to explicitly time-dependent problems in App.~\ref{timedependent}.\\

\noindent
Equation~\eqref{phsub} obeys the generic form
\begin{align}\label{eq:harmonicoscillator}
 i \partial_t \phi(q,t) = \frac{1}{2}\left[\hat{p}^2 + \omega^2\left(\hat{q} - \frac{\boldsymbol{\lambda}}{\omega}\cdot \textbf{R}(t)\right)^2 \right] \phi(q,t),
\end{align}
with a given initial state $\phi_0(q)=\phi(q,0)$, $q_0=\langle \phi_0 | \hat{q} |\phi_0 \rangle$ and $\dot{q}_0 = \langle \phi_0 | \hat{p} |\phi_0 \rangle$. As $\textbf{R}(t)$, i.e., the total dipole including nuclei, electrons and potentially external currents, is a given external perturbation in this context of the photonic subspace, the solution to the above equation is
\begin{align}\label{shiftcondition}
 \phi(q(t),t) = \hat{D}^\dagger(q(t))e^{-\frac{i}{2}\left[\hat{p}^2 + \omega^2\hat{q}^2 \right]t}\phi_0(q).
\end{align}
The coherent shift operator, which is a combination of a time-dependent translation and boost (translation in momentum space), fulfills condition~\eqref{shiftcondition} up to a time-dependent phase~\cite{merzbacher1998}
\begin{align}\label{coshifttd}
\hat{D}(q(t)) &= \exp{\left[-i\left(q(t) \hat{p} - \dot{q}(t) \hat{q}\right)\right]},
\end{align}
where the classical trajectory of the displacement coordinate $\hat{q}$ is given by
\begin{align*}
q(t) = &\int\limits_{0}^{t} dt' \sin\left[\omega(t-t')\right] \left( \boldsymbol\lambda \cdot \textbf{R}(t')\right)
\\ & \qquad + q_0 \cos(\omega t) + \dot{q}_0 \frac{\sin(\omega t)}{\omega}.
\end{align*}
We therefore see explicitly that a coherently driven photon mode just follows exactly the classical trajectory and only other observables provide access to the ``quantumness'' of the photon mode. That this holds is due to the quantization procedure of the electromagnetic field, which makes sure that without coupling to the matter subsystem the expectation values of the field operators reproduces the Maxwell equations in vacuum. So mode by mode the quantum harmonic oscillators need to reproduce the classical oscillators as long as we only have external sources~\cite{greiner1996, ruggenthaler2014}. In the time-independent case, where the classical equation of motion merely reduce to $\dot{q} =0$ and $q(t)= q_0$ we immediately arrive at the shifted eigenstates since it has to hold that $q_0=-\tfrac{\boldsymbol{\lambda}}{\omega}\cdot \textbf{R}$, where $\textbf{R}$ is the total static dipole. Consequently we have for the quantum states
\begin{align*}
 \hat{D}(q_0) \phi_k(q) = \phi_k(q - q_0),
\end{align*}
where we just used that $\hat{D}(q_0) \hat{q} \hat{D}^\dagger(q_0) = \hat{q} - q_0$. The resulting eigenenergies recover (due to the term $\tfrac{1}{2}(\tfrac{\boldsymbol{\lambda}}{\omega}\cdot \textbf{R})^2$) the original harmonic oscillator eigenenergies
\begin{align*}
\varepsilon_{k}(q_0) = \varepsilon(0) =  \omega \left(k + \frac{1}{2}\right) \, . 
\end{align*}
Since in the photonic subsystem we only have different shifted harmonic oscillators the resulting photonic wavefunction parametrically dependent on $\textbf{R}$ becomes
\begin{align*}
\Phi_k(\textbf{q},\{\textbf{R}\}) =   \prod_{\alpha}^{M_p} \phi_{\alpha,k_\alpha}(q_\alpha - q_\alpha^0)~,
\end{align*}
where we have defined $q_{\alpha}^{0} = -\tfrac{\boldsymbol{\lambda}_{\alpha}}{\omega_{\alpha}}\cdot \textbf{R}$ and we have a multi-index $k \equiv (k_0,...,k_{M_p})$ that collects the individual mode excitations, that is, a photonic multi-mode spectrum $\varepsilon_{k}$ is given by the energetic order of all possible photonic excitations. The low energetic polariton spectrum $E_\mu$ for very small $\omega_\alpha$ with weak coupling-strength is then for example dominated by replica of the ground-state with rising photonic occupation.  With this we find the parametrically-dependent photonic energy surfaces as
\begin{align}\label{eq:photonicsurface}
\varepsilon_{k}(\{\textbf{R}\}) = \sum\limits_{\alpha}^{M_p} \omega_\alpha \left(k_\alpha + \frac{1}{2}\right) \, .
\end{align}
As a consequence the photonic energy-surfaces in Eq.~\eqref{boham1} are just constants that merely shift the total energy. Therefore, the photons do not affect the coupled equations directly, since they do not enact a force on the other particles, that is, $\nabla_{R} \varepsilon_{k} = \textbf{0}$. Instead the photons affect the electrons and nuclei only via the non-adiabatic coupling elements. In contrast to the electron-nuclei coupling elements these photonic coupling elements are known analytically, that is,  given in Eq.~\eqref{couplings2} and \eqref{couplings3}. Thus the photons in the long-wavelength approximation do not introduce extra quantities that need to be determined numerically but rather change the usual Born-Huang expansion and lead to new but analytically known couplings.\\

\noindent
At this point we would like to note that the explicit solution of the photon subspace in terms of shifted harmonic oscillators corresponds to an adapted quantization procedure. Instead of quantizing the bare (zero-photon) vacuum, we quantize a non-zero (polarized) vacuum that corresponds to non-zero electromagnetic fields. This equivalence provides an interesting connection to trajectory-based approaches for matter-photon systems~\cite{hoffmann2018light}.

\section{Implications} \label{implications}
\noindent
After we have solved the photonic subsystem analytically in the presented Born-Huang expansion, let us restate the coupled problem that solves Eq.~\eqref{initialeigeneq} in a more compact form. By solving
\begin{align}\label{eq:electroniccompact}
&\left[\hat{H}_{e} + \hat{H}_{ne}(\textbf{r},\{\textbf{R}_n\}) + \sum\limits_{\alpha}^{M_p} \omega_\alpha \left(l_\alpha + \frac{1}{2}\right) \right] \psi^l_{\mu}(\textbf{r}, \{\textbf{R}_n\}) \notag\\
&- \frac{1}{2} \sum_{k=0}^{\infty} \left[ 2 \nabla^{lk} \cdot \sum_{j=1}^{N_e} \nabla_{\textbf{r}_j} + N_e \Delta^{lk} \right] \psi^k_{\mu}(\textbf{r}, \{\textbf{R}_n\}) \notag\\
&= E_\mu(\{\textbf{R}_n\}) \psi^l_{\mu}(\textbf{r}, \{\textbf{R}_n\})~,
\end{align}
and subsequently with the obtained polaritonic Born-Oppenheimer surfaces $E_\mu(\{\textbf{R}_n\})$
\begin{align}\label{eq:nuclearcompact}
&\left[ \hat{H}_{n} + E_\nu(\textbf{R}_n) \right] \chi^{\nu}_{i}(\textbf{R}_n) \notag\\
&+\sum_{\mu,l,k=0}^{\infty} \sum_{j=1}^{N_n} \frac{1}{2M_j} \bigg( Z_j\bigg[ \langle \psi^l_{\nu} \vert \psi^k_{\mu} \rangle_e \nabla^{lk} \cdot  \nabla_{\textbf{R}_j} \notag
\\
& \quad+ \langle \psi^l_{\nu} \vert \nabla_{\textbf{R}_j}\vert \psi^k_{\mu} \rangle_e \cdot \nabla^{lk} - \frac{Z_j}{2} \langle \psi^l_{\nu} \vert \psi^k_{\mu} \rangle_e \Delta^{lk}  \bigg]  \notag 
\\
&- \delta_{lk} \left[ 2 \langle \psi^l_{\nu} \vert \nabla_{\textbf{R}_j}  \vert \psi^l_{\mu} \rangle_e \!\!\cdot \!\! \nabla_{\textbf{R}_j}\!\! +\! \langle \psi^l_{\nu} \vert \Delta_{\textbf{R}_j}  \vert \psi^l_{\mu} \rangle_e \right]\!\bigg) \chi^{\mu}_{i}(\textbf{R}_n) \notag
\\
&= E_i \chi^{\nu}_{i}(\textbf{R}_n)~,
\end{align}
we find that the exact eigenstate is recovered as $\Psi_i(\textbf{R}_n, \textbf{r}, \textbf{q}) = \sum_{\mu,k=0}^{\infty} \chi_{i}^{\mu}(\textbf{R}_{n})\psi^{k}_{\mu}(\textbf{r}, \{ \textbf{R}_n \}) \Phi_k(\textbf{q}, \{\textbf{R} \})$. In the following, we consider a few implications of this new form of the original problem before discussing some approximation strategies.

\subsection{Relations to the Power-Zienau-Woolley transformation and the diabatic picture}\label{pzwequiv}
\noindent
Let us first compare the above form with the Power-Zienau-Woolley transformation~\cite{power1965introductory, andrews2018perspective} that allows to define a multipole form of the minimal coupling Hamiltonian. The length-form Hamiltonian that we use here can be obtained by approximating the quantized vector potential operator by its value at zero (or any other reference point) $\hat{\textbf{A}}(\hat{\textbf{r}}_i) \rightarrow \hat{\textbf{A}}(\textbf{0})$. This approximation leads to the momentum form of the minimal coupling Hamiltonian~\cite{tokatly2013,ruggenthaler2014, rokaj2017light} (without loss of generality we just show how the momentum of the electrons are adapted)
\begin{align}
\label{mincoup}
\frac{1}{2} \sum_{j=1}^{N_e} \left( -i\nabla_{\textbf{r}_j} - \sum_{\alpha=1}^{M_p} \frac{\boldsymbol\lambda_\alpha}{\omega_\alpha} \hat{p}_\alpha \right)^2,
\end{align}
where $\hat{p}_{\alpha}$ is as defined before.
Then performing an operator-valued boost of the form $ \exp [-i \hat{\textbf{R}} \cdot \sum_{\alpha=1}^{M_p} \frac{\boldsymbol\lambda_\alpha}{\omega_\alpha} \hat{p}_{\alpha}]$ leads to the length-form of Eq.~\eqref{hamiltonian}.
In the photonic subsystem in the Born-Huang expansion this operator-valued boost turns into a translation of the displacement coordinate for a given dipole moment $\textbf{R}$ and therefore is equivalent to $\hat{D}^\dagger(\textbf{q}_0)$, which is just the tensor product of the individual $\hat{D}^\dagger(q_0)$ as defined in Eq.~\eqref{coshifttd}. Thus when we apply the coherent shift operator in the Born-Huang expansion we take the step back to the momentum form of the long-wavelength approximated minimal-coupling Hamiltonian.\\
Formally we can connect the above operator-valued boost to a multipole expansion of the Power-Zienau-Woolley transformation $ e^{i \int d\textbf{r}\textbf{P}(\textbf{r}) \cdot \textbf{A}(\textbf{r})} $ of classical physics, where $\textbf{P}(\textbf{r})$ is a polarization field~\cite{goppert1931elementarakte,power1965introductory,craig1998}. By then promoting the classical quantities to operators a multipole form of the minimal-coupling Hamiltonian can be defined~\cite{goppert1931elementarakte,craig1998}. In the above momentum form of Eq.~\eqref{mincoup} it becomes straightforward to generalize beyond the long-wavelength approximation by reverting our very initial assumption of spatially independent modes and keeping explicitly $\boldsymbol{\lambda}_{\alpha}(\textbf{r})$~\cite{ruggenthaler2014, rokaj2017light}. Keeping the spatial dependence will change our expansion considerably, as it is no longer only the total dipole the photons couple to. In the general case the photon subsystem wavefunction will depend parametrically on all individual coordinates of the other particles, while in a formal multipole expansion higher order terms will be introduced. This will lead, similar to the decoupling of electrons and nuclei, to non-constant photonic surfaces that depend on the spatial position of the matter subsystem. As a consequence the combined system will try to occupy a minimal potential point given by the involved modes of the photon field.\\

\noindent
In the context of the momentum form of the non-relativistic QED Hamiltonian we also briefly want to highlight the difference between an adiabatic and diabatic picture~\cite{baer1992study}. In the length form we have decided to use the adiabatic picture, that is, we assume that the nuclei move on electronic surfaces and the electrons move on photonic surfaces. The physical rationale is that we assume that the nuclei ``move slower'' than the electrons, which in turn ``move slower'' (be aware that displacement coordinates $\textbf{q}$ are not real-space movements) than the photons. This has been expressed formally in our Born-Huang expansion by taking into account parametrically only the classical quantity $\textbf{R}$ for the photon subsystem wavefunction. But in principle we can use other arrangements. For instance, in Ref.~\cite{flick2017b} the photons were grouped with the nuclei due to their obvious similarity to the simplest approximations of the quantized nuclei by (harmonic) phonons. The electrons are therefore considered ``fast'' with respect to the rest. This is a diabatic picture. In this case it is the photons and nuclei that move on the electronic potential-energy surfaces.\\
Take, for example, the polaritonic wavefunction $\tilde{\Psi}_{\mu}$ but now we expand in terms of electronic subsystem wavefunctions parametrically dependent on the momentum-form photon coordinates and the position of the nuclei as
\begin{align*}
\tilde{\Psi}_{\mu}(\{\textbf{R}_n\}, \textbf{r},  \textbf{p}) = \sum_{k = 0}^{\infty} \psi_{k}(\textbf{r}, \{\textbf{R}_n, \textbf{p}\}) \Phi^k_{\mu}(\textbf{p},\{\textbf{R}_n\})~.
\end{align*}
If we consider the momentum form Eq.~\eqref{mincoup} we can get rid off the parametric photon-coordinate dependence by a boost in real space of the form $\exp[ -i \hat{\textbf{R}}\cdot \sum_{\alpha=1}^{M_p} \frac{\boldsymbol\lambda_\alpha}{\omega_\alpha} p_{\alpha}]$. This boost is the transformation from momentum to length form but with $p_{\alpha}$ as a number not as an operator. By this we have eliminated the dependence of the electronic wavefunction on the photon coordinate. Therefore, for what the photonic subsystem wavefunction $\Phi^k_{\mu}(\textbf{p},\{\textbf{R}_n\})$ is concerned, the electronic potential-energy surfaces become flat and it is only coupling elements that will change the photon field. In the diabatic case, however, there are usually no analytic solutions for the electrons available.\\

\noindent
So far, we have focused on how the standard Born-Huang expansion is changed due to analytically known non-adiabatic coupling elements introduced by the photons. Alternatively one could instead of these coupling elements include parts of the matter-photon interaction directly in the matter subsystems, for example, in the diabatic picture we could include $ q_{\alpha} \boldsymbol\lambda_\alpha \cdot \hat{\textbf{R}} $ and $(\boldsymbol\lambda_\alpha \cdot \hat{\textbf{R}})^2$ in the electronic equation and vary parametrically $q_{\alpha}$. The resulting cavity-adapted electronic eigenstates will already incorporate certain effects of the coupling (see Ref.~\cite{flick2017b}). This alone can already lead to a relatively accurate treatment of ground and dark states. However, while this cavity Born-Oppenheimer approximation is accurate for energetically well-separated eigenstates, it does usually not capture the hallmark of strong coupling, that is, the polariton formation due to the Rabi splitting (see for example table 1 in Ref.~\cite{flick2017b}). For the experimentally easily accessible coupling values and on resonance with the bare electronic transition, the upper and lower polariton states are energetically close and without incorporating the non-adiabatic couplings the theoretical description is not accurate enough. This changes for ultra-strong coupling, since the resulting parametrically shifted surfaces become then energetically well separated. While from a quantum-optical perspective it is usually the excited-state structure that is important, for certain chemical aspects the coupled matter-photon ground state might be sufficient. Especially in the latter case the cavity Born-Oppenheimer approach of Ref.~\cite{flick2017b} represents an interesting alternative. The cavity Born-Oppenheimer approach then needs an adjustment of common quantum chemical codes by including the $(\boldsymbol\lambda_\alpha \cdot \hat{\textbf{R}})^2$ and $ q_{\alpha} \boldsymbol\lambda_\alpha \cdot \hat{\textbf{R}}$ parts, with a potentially large parametrically scan of the photonic displacements. The primary benefit of this method is highlighted in the discussion of Sec.~\ref{2dgaas}.\\

\noindent
Let us at this point also highlight, that the length and momentum form of the non-relativistic QED Hamiltonian in the long-wavelength approximation, get their name from the above discussed translations and boosts. In the momentum form, a translation in real space has no effect on the photons at all, and a boost of the matter subsystem merely shifts the photon field coherently, leaving the eigenstates invariant. In the momentum form it is the variations/fluctuations in momentum space, for example, due to non-zero momentum matrix elements $\nabla^{lk}$, that lead to physical effect on the eigenstates. In the length form (after we have performed already an operator-valued boost) we can translate the matter subsystem in real space and it only amounts to a simple coherent shift of the photonic subsystem, which again leaves the eigenstates invariant. It is the real-space variations/fluctuations, for example, due to non-zero dipole matrix elements, that lead to physical effects on the eigenstates (see Sec.~\ref{modelsystem} for an example).

\subsection{The relevance of self-polarization}\label{relselfpol}
\noindent
The physical and mathematical relevance of the self-polarization, also called the dipole self-energy, $\sum_\alpha (\boldsymbol\lambda_\alpha \cdot \hat{\textbf{R}})^2/2 $ has been discussed in multiple publications~\cite{vukics2015,pellegrini2015,flick2017ab,rokaj2017light,cs2018entangled}. The here presented reformulation of the non-relativistic QED problem in the long-wavelength approximation allows an alternative, very intuitive and physical explanation of the most important fact, that is, that the coupled nucleus-electron-photon system is not stable without the self-polarization~\cite{rokaj2017light}. This becomes evident if one considers the photonic surfaces $\varepsilon_k$ as defined in Eq.~\eqref{eq:photonicsurface}. In the case that the self-polarization term is discarded in the original Eq.~\eqref{eq:harmonicoscillator}, the energy expressions $\varepsilon_{k}$ get a dipole-dependent shift $-(\boldsymbol{\lambda}\cdot\textbf{R})^2/2$. Therefore the photonic potential-energy surfaces $\varepsilon_l$ in Eq.~\eqref{eq:electroniccompact} would not be constant, instead they would be quadratically divergent. This leads to a linearly divergent force 
\begin{align*}
\nabla_R \varepsilon_{l}^{\text{no R}^2} (\{\textbf{R}\}) = -\sum_\alpha \boldsymbol\lambda_\alpha \left(\boldsymbol \lambda_\alpha \cdot \textbf{R}\right)~.
\end{align*}
As a consequence Eq.~\eqref{eq:electroniccompact} does not posses an equilibrium solution (if we consider infinite space), i.e, $E_{\mu}(\{\textbf{R}_n\}) \rightarrow -\infty$, and with this also the original problem becomes unbounded from below, that is, $E_{i} \rightarrow -\infty$. If we restrict thus to a finite simulation box (which is not connected to the much larger quantization volume or cavity volume) then the ratio between coupling strength and box size (which in turn restricts the value of $\textbf{R}$) determines whether we get a stable system or whether the minimal energy is reached by tearing the system apart (nuclei on one side and electrons on the other to maximize $\textbf{R}$). With the self-polarization such a problem does not appear and a stable, simulation-box-size independent solution is well-defined. That this problem is not more often encountered has to do with the fact that mostly the coupled system is solved in a restricted state space. As long as our electronic expansion  is small (see the discussion in Secs.~\ref{roothaanhall}, \ref{modelsystem} and \ref{numchecks}) this divergence does not emerge because the limited set of electronic states cannot describe such a divergent shift. Take, for instance, the minimal setting of only a single electronic excitation. The system is then represented by the Pauli-matrix algebra with $\hat{\textbf{R}} \rightarrow \hat{\sigma}_z$ and $(\boldsymbol\lambda_\alpha \cdot \hat{\textbf{R}})^2 \rightarrow \lambda_\alpha^2 \hat{\mathds{1}}$ with the identity on the two dimensional vector space denoted by $\hat{\mathds{1}}$. The self-polarization has no effect in this smallest electronic space besides a shift in energy. However, this is not the case for the exact solution and the missing contribution is essential to capture ground-state observables. This effect will be illustrated by a numerical example in Sec.~\ref{numchecks}.\\

\noindent
Let us finally remark that the self-polarization contribution is necessary to connect the momentum and the length form of the non-relativistic QED Hamiltonian in the long-wavelength approximation and that the self-polarization term is not equivalent to $\hat{\textbf{A}}^{2}(0)$. Further the influence of the self-polarization term for dynamical phenomena and for weak to moderate light-matter interaction strength is less evident. Only after sufficiently long propagation large deviations will be observable. 

\subsection{Observables}\label{phtonic_obs}
\noindent
Although excitations of the photonic system are just implicitly addressed in the nuclear \eqref{eq:nuclearcompact} and electronic \eqref{eq:electroniccompact} components, observables can be calculated as usual
\begin{align*}
&\langle \Psi_i \vert \hat{O}\vert \Psi_i \rangle
= \langle \sum_{\mu,l = 0}^{\infty} \chi^{\mu}_{i}  \psi^{l}_{\mu} \Phi_{l} \vert \hat{O} \vert \sum_{\nu, k = 0}^{\infty} \chi^{\nu}_{i}  \psi^k_{\nu}\Phi_{k}\rangle~.
\end{align*}
For instance, we can consider photonic observables, such as the mode-occupation. The physical interpretation of displacement $\hat{q}_{\alpha}$ and furthermore creation and destruction operators change between momentum and length form, that is, the occupation of mode $\alpha$ in momentum form is related to the electro-magnetic field energy-density~\cite{craig1998}. We solve the photonic system as discussed in Sec.~\ref{photonsub}. The coherent translation $\hat{D}$ transfers the wavefunction $\Psi_i \rightarrow \tilde{\Psi}_i$ into the momentum frame while the operators remain formally identical. In our current formulation this leads to the following mode-resolved form
\begin{align*}
&\langle \tilde{\Psi}_i \vert \hat{\mathbb{1}}_n \otimes \hat{\mathbb{1}}_e \otimes \hat{a}^\dagger_\alpha \hat{a}_\alpha \vert \tilde{\Psi}_i \rangle\\
&= \sum_{\mu, \nu, l,k = 0}^{\infty} \langle \chi^{\mu}_{i} \vert \chi^{\nu}_{i} \rangle_n \langle \psi^l_{\mu}(\{\textbf{R}_n\}) \vert \psi^k_{\nu}(\{\textbf{R}_n\}) \rangle_e\\ &\times \langle \Phi_{l}(\{\textbf{R}\}) \vert \hat{a}^\dagger_\alpha \hat{a}_\alpha \vert \Phi_{k}(\{\textbf{R}\})\rangle_p\\
&= \sum_{\mu, \nu, l = 0}^{\infty} \langle \chi^{\mu}_{i} \vert \chi^{\nu}_{i} \rangle_n \langle \psi^l_{\mu}(\{\textbf{R}_n\}) \vert \psi^l_{\nu}(\{\textbf{R}_n\}) \rangle_e l_\alpha ~.
\end{align*}
In this way we can construct similar photonic observables such as $\hat{q}_{\alpha}$ and any other observable of the combined nucleus-electron-photon system. But of course, we can also access other observables, such as the electronic density or excitation energies (see Sec.~\ref{numbench} for examples). As it will turn out, depending on which observable we want to consider, we will need different levels of approximations to the full Born-Huang expansion to get accurate results. Overly reduced models will not be able to capture - even qualitatively - some considered observables.

\subsection{Born-Oppenheimer including photons: the explicit polariton approximation}\label{nboa}
\noindent
The coupled equations of the Born-Huang expansion, that is, Eqs.~\eqref{eq:electroniccompact} and \eqref{eq:nuclearcompact} in its compact form, can be drastically simplified if certain coupling elements can be neglected, that is, for specific subsystem wavefunctions the derivative with respect to their parametrical dependence vanishes. This is equivalent to the fact that the induced energetic surfaces are well separated. That this holds true for all electronic subsystem wavefunctions is one of the basic assumptions of the traditional Born-Oppenheimer approximation of quantum chemistry. If we make the same assumption we still have the photonic coupling elements that connect the different nuclear subsystem wavefunctions. So in order to arrive at a similar simple form as the original Born-Oppenheimer approximation we have to also assume that these coupling elements are negligible. This will be the case if we - besides the usual physical rationale of ``slowly moving'' (almost classical) nuclei - also assume that the frequencies of the photonic modes is mainly in the electronic energy range. In this case it seems reasonable to assume that the coupling between nuclei due to the photons is not important, and we only consider photons that couple to the electronic system. This is similar to the matter-photon coupling in Ref.~\cite{galego2015}, that only considers the electronic dipole contribution. Under those assumptions, electron-photon and nuclear coordinates factorize, leading to
\begin{align*}
\Psi_{i} \approx \Psi_{\nu}^{\mu}(\textbf{R}_n, \textbf{r}, \textbf{q}) = \chi^{\mu}_{\nu}(\textbf{R}_n) \sum_{k = 0}^{\infty}  \psi^k_{\nu}(\textbf{r},\{\textbf{R}_n\}) \Phi_{k}(\textbf{q},\{\textbf{R}\})~.
\end{align*}
The many-body wavefunction $\Psi_{i}$ is now approximated by a polaritonic excitation $\nu$ and uncorrelated nuclear vibrations $\mu$ with ground-state $\Psi_{0}^{0}(\textbf{R}_n, \textbf{r}, \textbf{q}) = \chi^{0}_{0}(\textbf{R}_n) \sum_{k = 0}^{\infty}  \psi^k_{0}(\textbf{r},\{\textbf{R}_n\}) \Phi_{k}(\textbf{q},\{\textbf{R}\})$ . As a consequence of the Born-Oppenheimer approximation, the nuclear excitations are calculated on polaritonic quasi-particle energy-surfaces.
The photonic contribution depends parametrically on the total dipole $\textbf{R}$. 
This leads to the following simplified form of the Born-Huang expansion
\begin{align}\label{maineq}
\begin{split}
E_{\nu}&(\{\textbf{R}_n\})  \psi^l_{\nu}(\textbf{r},\{\textbf{R}_n\}) =  \hat{H}_{BO}^l \psi^l_{\nu}(\textbf{r},\{\textbf{R}_n\}) \\
&-\sum_{k=0}^\infty \left[ \nabla^{lk} \cdot \sum\limits_{i=1}^{N_e} \nabla_{\textbf{r}_i} + \frac{N_e}{2} \Delta^{lk} \right] \psi^k_{\nu}(\textbf{r},\{\textbf{R}_n\})~,
\end{split}
\end{align}
where the electron-photon-solution feeds into the nuclear equation
\begin{align}\label{nuclBO}
\left(-\frac{1}{2}\sum_{j=1}^{N_n} \frac{1}{M_j} \nabla^2_{\textbf{R}_j} + E_{\nu}(\textbf{R}_n)\right) \chi^{\mu}_{\nu}(\textbf{R}_n) = E_{\nu}^\mu \chi^{\mu}_{\nu}(\textbf{R}_n)~.
\end{align}
Clearly, other approximations of the fully coupled Born-Huang expansions are possible, for example, we could assume that the photonic modes are only within the nuclear-excitation energy range. To distinguish the above specific choice we call the coupled Eqs.~\eqref{maineq} and \eqref{nuclBO} the \textit{explicit polariton approximation} to highlight that we consider nuclei moving on polaritonic surfaces.\\

\noindent
To solve the explicit polariton equations we restrict first our photonic excitation space to a maximum number of possible excitations $ l \in \{0,1,..., l_{max}\} $, that is, the sum $\sum_{k=0}^\infty \rightarrow \sum_{k=0}^{l_{max}}$ is truncated at a finite photonic occupation.  Expanding the full basis explicitly corresponds to an expansion in generalized coherent eigenstates. The cost we then pay is that we effectively end up with the same high-dimensional problem of the initial fully coupled electron-photon Hamiltonian in a restricted photon-space. For the electronic subsystem we will in the following, however, also use a suitable basis expansion which we will truncate in practice. How this is done and how the polaritonic problem of Eq.~\eqref{maineq} is expressed in such a basis expansion we will discuss in Sec.~\ref{roothaanhall}. The explicit polariton in a finite basis expansion (for photonic and electronic subsystem, respectively) can be efficiently used to solve coupled systems for weak and strong coupling for finite and also periodic systems (see App.~\ref{periodic}), where the interpretation of arising polaritonic bands does conceptually not deviate from the finite molecular problem. The additional effort to include the photonic interaction into a common solution of the nuclear-electron problem is computationally negligible as long as we can treat the relevant set of the excited electronic states efficiently and the accuracy of the restricted electronic and photonic basis can be estimated based on physical intuition or arguments. That this is not always as intuitive as one might hope for and that there is a subtle interplay between number of electronic and photonic excitations that we take into account we will show explicitly in Sec.~\ref{numchecks}. Further, while certain simplified quantities such as changes in excitation energies due to the emergence of polaritonic states might seem already converged only after taking into account a minimal number of photonic excitations and electronic basis states, the convergence in other quantities and especially of the underlying approximate wavefunction with respect to the photonic and electronic basis set might be much slower. We will investigate the restricted basis issue later and will as reference use the frequently employed restriction to merely just one photonic excitation per mode. We will name the resulting approximate solution the \textit{single photon polariton} that involves at maximum a single excitation per mode $l_\alpha \in \{0,1\}$. In Sec.~\ref{modelsystem} we will use this approximation to connect the above quantum-chemical approach to quantum-optical models that are used to describe coupled matter-photon systems in a simplified way and we will discuss implications.\\

\noindent
While the above explicit polariton approximation is expected to be a reliable approximation for the weak and strong coupling regime within a restricted set of excitations, in the domain of ultra-strong coupling, that is, $\lambda_\alpha/\sqrt{2\omega_\alpha} \sim 1$, other possible Born-Huang expansions, for example, such as the diabatic approach briefly discussed in Sec.~\ref{pzwequiv}, might be more appropriate.

\subsection{Structural similarity to Floquet theory}\label{floquet}
\noindent
Let us stress the connection of the explicit polariton approximation and the Floquet approach.
In Floquet theory, the Hamiltonian is time-periodic 
$\hat{H}(t)=\hat{H}(t+T)$ 
with the period $T=\frac{2\pi}{\omega}$. Most commonly, this is achieved by driving a time-independent system $\hat{H}_s$ with a periodic external driving $\hat{D}(t) = \hat{\boldsymbol r}\cdot \textbf{E}(t)$, where often a monochromatic field is applied $\textbf{E}(t) = \textbf{E}_0 \cos(\omega t)$. More general, the full minimal coupling 
\begin{align*}
-\frac{1}{2} \sum_{i=1}^{N_e} \nabla_i^2 - \textbf{A}(t)\cdot \sum_{i=1}^{N_e} (-i\nabla_i) + \frac{N_e}{2} \textbf{A}^2(t) 
\end{align*}
should be considered. The driving vector-potential resembles then for example $\textbf{A}(t) = \textbf{A}_0 \sin(\omega t)$. The solution to the time-periodic Sch\"odinger equation can be achieved by the Floquet-Bloch ansatz
\begin{align*}
\Psi_\mu(\textbf{r}t) = \sum_{\nu=1}^{\infty} c_{\nu \mu} e^{-i\varepsilon_\nu t} \sum_{n=-\infty}^{\infty} e^{-in\omega t}\phi_n^\nu (\textbf{r})~.
\end{align*}
While the Hamiltonian has to satisfy the periodicity condition, the wave-function can be of any arbitrary period, including for example Rabi-oscillations, as a consequence of linear combinations $c_{\nu \mu}$. A numerically feasible solution can then be achieved by solving the Floquet-matrix equation
\begin{align*}
\mathcal{H}^{mn} \phi_n^\nu (\textbf{r}) =  \varepsilon_\nu \phi_m^\nu (\textbf{r})
\end{align*}
with
\begin{align*}
\mathcal{H}^{mn} = \frac{1}{T} \int\limits_{0}^{T} dt e^{i(m-n)\omega t} H(t) + \delta_{mn} m \omega \mathbb{1}
\end{align*}
in a restricted subset of possible excitations $m,n \in \{-n_{max}, ..., -1, 0, +1, ..., n_{max}\}$. The driving then leads in linear order to $\frac{1}{T} \int_{0}^{T} dt e^{i(m-n)\omega t} \textbf{A}(t) \sim i \delta_{m,n\pm1}$ and in second order, that is, $\frac{1}{T} \int_{0}^{T} dt e^{i(m-n)\omega t}\textbf{A}^2(t) \sim  \delta_{m,n\pm\{0,2\}}$ to connections that resemble the non-adiabatic coupling elements $\nabla^{lk}, ~ \Delta^{lk}$ of equation \eqref{maineq}. The according eigenvectors $\phi^\nu (\textbf{r}) = (...\phi_{-1}^\nu (\textbf{r}) \phi_{0}^\nu (\textbf{r}) \phi_{+1}^\nu (\textbf{r}) ...)^T$ posses the structure of differently weighted electronic solutions associated with an excitation or ``photon''-number $n$. While it is intuitive to consider quantized excitations due to a classical electromagnetic field as photons, the rigor of this interpretation of course relies on the similarity to the QED formulation.\\

\noindent
As we have seen in equilibrium, the effect of the photons on the electronic structure is purely determined by fluctuations within the shifted harmonic oscillators which connect electronic excitations. The classical contribution, namely the shift $\hat{q}_\alpha \rightarrow \hat{q}_\alpha - q_\alpha^0$, is without effect on the electronic structure.
This clarifies that the eigenstates of Eq.~\eqref{maineq} are polaritonic wavefunctions projected on the electronic subspace, that is, the solution of Eq.~\eqref{maineq} can be constructed as a vectorial expansion for each mode in sub-blocks according to $ \psi_{\nu} = \{ \psi^0_{\nu},\psi^1_{\nu}, ..., \psi^{l_{max}}_{\nu}\} $. The explicit polariton is structurally very similar to solving the Floquet problem as defined above. Two essential differences arise, that is, the energy is bounded from below $k \geq 0$ and the couplings $\nabla^{lk},~\Delta^{lk}$ obey a $\sqrt{k}$ scaling which coincides with the classical Floquet description for large photon-numbers $k\rightarrow \infty$ and coherent (classical) field-states as pointed out by Ref.~\cite{shirley1965solution,gao2000nonperturbative}. As a direct consequence of the lower bound, all occupations are well known in contrast to Floquet theory. While in Floquet theory the $n=0$ sector is not distinct from other components, that is, the solution is even invariant under arbitrary shifts $\varepsilon'_\nu = \varepsilon_\nu + n\omega,~n \in \mathbb{Z}$, in QED every photonic sector, especially the ground-state, is unique. Therefore Eq.~\eqref{maineq} shows how the typical Floquet equation has to be adjusted to adhere to QED, also avoiding usual drawbacks of Floquet theory.\\

\noindent
Furthermore, the structural similarity between QED and Floquet theory (while the physical interpretation and experimental set-ups are very different) allows to adjust currently available Floquet-implementations to be able to consider QED-chemistry situations. Physically we exchange a classical external driving with a large number of photons by a quantized photonic field considering typically small photon numbers and vice versa. Furthermore, this means that a vast amount of novel Floquet-engineered effects~\cite{hubener2017} can be qualitatively reproduced with equilibrium QED, where no external driving is necessary. This is especially interesting since the usual drawbacks of external driving, for example, heating, can be avoided, which might prove helpful in stabilizing Floquet-engineered states. Our derivations furthermore present the necessary generalization to the full nucleus-electron-photon problem.

\section{Connection to quantum-optical models}\label{connection_qo}
\noindent
We have now approximated the full Born-Huang expansion by the explicit polariton approximation given in Eqs.~\eqref{maineq} and \eqref{nuclBO}. This has already reduced the complexity considerably. Still solving the explicit polariton approximation in full real space for the electrons is very challenging due to the involved Coulomb interaction among the particles. Thus for practical purposes also a restriction of the electronic basis set is desirable. Also in most models for coupled matter-photon systems the solution of the full many-electron problem is avoided by working in a restricted basis for the electrons. In its simplest form this restriction leads to a two-level approximation to the matter subsystem, which is a common approximation strategy encountered in quantum optics, for example, in the Jaynes-Cummings model or for many two-level systems in the Tavis-Cummings model. In order to connect to these well-known models and highlight possible problems in an overly simplified treatment as well as for practical purposes we will in the following also restrict the electronic basis set.

\subsection{Basis-expansion in electronic eigenstates}\label{roothaanhall}
\noindent
To restrict also the electronic subspace we expand the polariton in a common electronic basis. Although this can be any complete basis, the exact electronic eigenfunctions will simplify the following steps. What will emerge is a way how to use the basis such that the photonic influence can be treated by a simple diagonalization in this basis. As a (not necessarily orthonormal) basis we choose some electronic wavefunctions $\varphi_n(\textbf{r}, \{\textbf{R}_n\})$ and then we expand the many-body electronic wavefunction 
\begin{align*}
\psi^l_{\nu}(\textbf{r},\{\textbf{R}_n\}) = \sum_{n=0}^\infty c^{\nu}_{nl}(\{\textbf{R}_n\}) \varphi_n(\textbf{r},\{\textbf{R}_n\})~.
\end{align*}
This expansion applied to Eq.~\eqref{maineq} leads to
\begin{align}\label{rothaan_hall}
\begin{split}
E_\nu &\sum_{n=0}^\infty S^{jn} c^\nu_{nl}  =  \sum_{n=0}^\infty H^l_{BO,jn} c^\nu_{nl}\\ &- \sum_{n,k=0}^\infty \bigg[ (-\nabla_{e}^{jn} c^\nu_{nk})\cdot \nabla^{kl} + \frac{N_e}{2} S^{jn} c^\nu_{nk} \Delta^{kl} \bigg],
\end{split}
\end{align}
where
\begin{align*}
S^{jn} &= \langle\varphi_j \vert \varphi_n \rangle_e\\
H^l_{BO,jn} &= \langle\varphi_j \vert \hat{H}^l_{BO} \vert \varphi_n \rangle_e\\
\nabla_{e}^{jn} &= \langle\varphi_j \vert \sum_{i=1}^{N_e}\nabla_{\textbf{r}_i} \vert \varphi_n \rangle_e\\
1 &= \sum_{l=0}^{\infty}\sum_{n,n'=0}^\infty c^{\nu}_{ln} S^{nn'} c^{\nu}_{n'l}  ~.
\end{align*}
Here we suppressed the parametric dependence on the nuclear configuration for brevity and we did not assume that the basis is orthonormal. We remind the reader that the analytically known coupling elements $\nabla^{kl}$ and $\Delta^{kl}$ are completely independent of nuclear and electronic coordinates.\\

\noindent
Along this line it becomes evident how the non-adiabatic nuclear-polariton coupling elements $\langle \psi^l_{\nu} \vert \nabla_{\textbf{R}_j}\vert \psi^k_{\mu} \rangle_e$ can be calculated from the knowledge of bare coupling elements $\langle \varphi_n \vert \nabla_{\textbf{R}_j}\vert \varphi_m \rangle_e$. We expand into
\begin{align}\label{nonadiabatic_elements_nuclpol}
&\langle \psi^l_{\nu} \vert \nabla_{\textbf{R}_j}\vert \psi^k_{\mu} \rangle_e = \sum_{n,m=0}^\infty c^{\nu}_{ln}(\{\textbf{R}_n\}) \bigg[ S^{nm} \notag \nabla_{\textbf{R}_j}c^{\mu}_{mk}(\{\textbf{R}_n\}) \\
&+  c^{\mu}_{mk}(\{\textbf{R}_n\}) \langle \varphi_n(\textbf{r},\{\textbf{R}_n\}) \vert \nabla_{\textbf{R}_j} \vert \varphi_m(\textbf{r},\{\textbf{R}_n\}) \rangle_e\bigg]
\end{align}
and observe that two contributions occur. The second line represents the superposition of bare non-adiabatic elements according to the participation of those states in the polariton solution. Furthermore, the linear coefficients themselves introduce an additional component $\nabla_{\textbf{R}_j}c^{\mu}_{mk}(\{\textbf{R}_n\})$ as their value depends for example on the dipole-transition element $\langle \varphi_n(\textbf{r},\{\textbf{R}_n\}) \vert \hat{\textbf{r}}_j \vert \varphi_m(\textbf{r},\{\textbf{R}_n\}) \rangle_e$ which will change depending on the nuclear configuration as discussed in Sec.~\ref{modelsystem}.\\

\noindent
Next we choose a basis and here we draw the connection to quantum chemistry by a selection of Slater determinants. For this, we multiply from the left with an electronic (excited) Slater determinant 
\begin{align*}
&\varphi_n(\textbf{r},\{\textbf{R}_n\}) = \hat{T}^{(n)}\varphi_0(\textbf{r},\{\textbf{R}_n\})\\
&= \hat{T}^{(n)} \frac{1}{\sqrt{N_e}}
\begin{vmatrix}
\varphi_1(r_1,\{\textbf{R}_n\}) & ... & \varphi_1(r_{N_e},\{\textbf{R}_n\})\\
... & ... & ...\\
\varphi_{N_e}(r_1,\{\textbf{R}_n\}) & ... & \varphi_{N_e}(r_{N_e},\{\textbf{R}_n\})
\end{vmatrix}_{-}
\end{align*}
with single particle orbitals $\varphi_j(r_{m},\{\textbf{R}_n\})$ and the operator for the n-th excitation $\hat{T}^{(n)}$. The operator $\hat{T}^{(n)}$ excites the wavefunction n-times, that is, $\langle \varphi_i(\textbf{r},\{\textbf{R}_n\}) \vert \hat{T}^{(j)}\varphi_0(\textbf{r},\{\textbf{R}_n\})\rangle_e = S^{ij}$. Further, within Hartree-Fock theory, the photonic coupling elements would open the usually closed Roothaan-Hall equation~\cite{von1998encyclopedia} into all possible excited components. Here, however, we do not restrict to Hartree-Fock theory but rather perform a configuration-interaction expansion of Eq.~\eqref{rothaan_hall} within a restricted subspace of photonic and electronic excitations or with a variational minimization of the polaritonic energy according to
\begin{align*}
E_\nu &= \sum_{l,j=0}^{\infty}\sum_{n=0}^\infty c^{\nu}_{lj} H^l_{BO,jn} c^\nu_{nl}\\ &- \sum_{l,j=0}^{\infty} \sum_{n,k=0}^\infty c^{\nu}_{lj} \bigg[ (-\nabla_{e}^{jn} c^\nu_{nk})\cdot \nabla^{kl} + \frac{N_e}{2} S^{jn} c^\nu_{nk} \Delta^{kl} \bigg]~.\notag
\end{align*}
In this form it becomes apparent that we can in principle use and extend common quantum chemical minimization procedures, such as Configuration-Interaction or Coupled-Cluster techniques~\cite{purvis1982full}, to solve the polaritonic problem.

\subsection{A natural connection to model-systems}\label{modelsystem}
\noindent
Let us now build the connection to a very common system applied in quantum optics. We assume a set of identical, independent molecules, the excited Slater-determinants are then exact solutions of the electronic system, coupled by one photonic mode $\alpha=1$. With close to degenerate energies, the first excited component of this single photon polariton subspace is represented by the unperturbed set
\begin{align*}
&\{ \psi^{l=1, bare}_{\nu=1} = \varphi^{MB}_{n=0} ~;~E_{\nu=1} = E_0 + \omega\} \\ &\{\psi^{l=0,bare}_{\nu=1} = \varphi^{MB}_{n=1}~;~ E_{\nu=1} = E_0 + \Delta E_e\}
\end{align*}
with bare electronic excitation $\Delta E_e$ and detuning of the photon mode $\delta$
\begin{align*}
\Delta E_e = E_1-E_0~, \quad \delta = \omega - \Delta E_e~.
\end{align*}
Hereby
\begin{align*}
\varphi^{MB}_{n=1} = \hat{S}^{+} \hat{T}^{(1)} \hat{S}^{-}\varphi_{0}^{1} \otimes \varphi_{0}^{2} \otimes ... \otimes \varphi_{0}^{N}
\end{align*}
is an anti-symmetric $(\hat{S}^-)$ many-body fermionic wavefunction with a symmetrized $(\hat{S}^+)$ excitation represented by the single excitation operator $\hat{T}^{(1)}$. In addition to this fully symmtrized excitation, there exist $N_e-1$ excitations with mixed symmetry which are often referred to as dark states. They posses a vanishing transition dipole due to anti-symmetric components in the many-body excitation~\cite{galego2015} and do not directly couple to the transversal mode by emission or absorption of a single photon. Nevertheless, higher order processes such as the Lamb shift still affect those states. As is usually done, we will disregard these dark states in the following. Next, instead of the electronic momentum elements $\nabla_e^{jn}$, in molecular systems the dipole moment is typically preferred in calculations . It is connected to the momentum element $\nabla_e^{jn}$ by the operator identity
\begin{align*}
\begin{split}
[\hat{H},\hat{\textbf{r}}]_{-} &= \left[ \frac{1}{2} \hat{\textbf{p}}^2 + \hat{W}_{ee}(\textbf{r},\textbf{r}') + \hat{H}_{ne}(\textbf{r}), \hat{\textbf{r}} \right]_{-}= -i\hat{\textbf{p}}\\ &= - \nabla_\textbf{r}~.
\end{split}
\end{align*}
In the approximation of electronically non-interacting molecules, that is $ \hat{W}_{ee}(\textbf{r},\textbf{r}') = 0 $, the electronic eigenvalues are isolated molecular excitations
\begin{align*}
\nabla_{e}^{jn} = - \langle j \vert [\hat{H},\hat{\textbf{r}}]_{-} \vert n \rangle_e = (E_n-E_j) \langle j \vert \hat{\textbf{r}} \vert n \rangle_e = \Delta E_{nj}\textbf{r}_{jn}
\end{align*}
and the energy-difference leads to a transformation of the characteristic bilinear coupling scale
\begin{align*}
\frac{1}{\sqrt{2\omega}} \boldsymbol\lambda \cdot \nabla_{e}^{jn} &= \frac{E_n-E_j}{\sqrt{2\omega}} \boldsymbol\lambda \cdot \textbf{r}_{jn}\\ \rightarrow \frac{\Delta E_e}{\sqrt{2\omega}} \boldsymbol\lambda \cdot \textbf{r}_{01} &= \frac{\omega - \delta}{\sqrt{2\omega}} \boldsymbol\lambda \cdot \textbf{r}_{01}~.
\end{align*}
The coupling elements in length and momentum form are identical up to a factor $(\omega-\delta)/\omega$~\cite{craig1998}.
The position $\hat{\textbf{r}}$ is defined in relation to the center-of-charge.
We now solve this ansatz in the smallest possible subspace for electronic $j$ and photonic excitations $l$ such that $(j,l) \in \{(0,0),(0,1),(1,0),(1,1)\}$ were we choose the pure electronic Slater-determinants $\varphi^{MB}_n$ to form an orthonormal basis
\begin{align*}
S^{jn} = \delta_{jn}~.
\end{align*}
The two-level single photon polariton approximation becomes extensively problematic as coupling or detuning increases and is not suited to describe the ground-state of a correlated system although it includes the anti-rotating contributions, that is, there is no rotating frame assumed. Especially the self-polarization contribution can become troublesome as discussed in Sec.~\ref{relselfpol}. This is made more explicit and elaborated in Sec.~\ref{numchecks}.

The single photon polariton with a single electronic excitation can be represented by the 4-by-4 matrix
 \begin{widetext}
\begin{align*}
\left(
\begin{matrix}
 H_{BO,00}^{l=0} - \frac{N_e}{2} \Delta^{00} & 0 & 0 & \nabla_{e}^{01} \cdot\nabla^{10} \\ 
0 & H_{BO,00}^{l=1} - \frac{N_e}{2} \Delta^{11} & \nabla_{e}^{01} \cdot\nabla^{01} & 0 \\
0 & \nabla_{e}^{10} \cdot\nabla^{10} & H_{BO,11}^{l=0} - \frac{N_e}{2} \Delta^{00} & 0 \\
\nabla_{e}^{10} \cdot\nabla^{01} & 0 & 0 & H_{BO,11}^{l=1} - \frac{N_e}{2} \Delta^{11}
\end{matrix}
\right)
\end{align*}
\end{widetext}
which can be block diagonalized in a degenerate single excitation 2-by-2 matrix subspace 
\begin{align*}
\left(
\begin{matrix}
H_{BO,00}^{l=1} - \frac{N_e}{2} \Delta^{11} - E_1 & \nabla_{e}^{01}\cdot \nabla^{01} \\
\nabla_{e}^{10}\cdot \nabla^{10} & H_{BO,11}^{l=0} - \frac{N_e}{2} \Delta^{00} - E_1
\end{matrix}
\right)
\cdot
\left(
\begin{matrix}
c_{01}\\
c_{10}
\end{matrix}
\right)
\end{align*}
whose solution resembles the upper and lower polariton with an energy- or Rabi-splitting of
\begin{align*}
\Delta E = \sqrt{ \tilde{\delta}^2 + \tilde{\Omega}^2 }~,
\end{align*}
the generalized collective detuning
\begin{align*}
\tilde{\delta} = \delta + \mathcal{L}
\end{align*}
and the collective Rabi-frequency
\begin{align*}
\tilde{\Omega} &= \sqrt{N_e} 2\frac{\omega - \delta}{\sqrt{2\omega}}\vert \boldsymbol\lambda \cdot  \textbf{r}_{01}^{\{\textbf{R}_n\}} \vert\\  \textbf{r}_{01}^{\{\textbf{R}_n\}} &= \langle \varphi_{n=0}^{1}(\{\textbf{R}_n\}) \vert \hat{\textbf{r}} \vert \varphi_{n=1}^{1}(\{\textbf{R}_n\}) \rangle_e~.
\end{align*}
The conserving fluctuations $\Delta^{ll}$ introduce a collective frequency or diamagnetic shift
\begin{align}\label{lambshift}
\mathcal{L} = \left[\sqrt{N_e} \frac{\boldsymbol\lambda}{\sqrt{2\omega}} \right]^2
\end{align}
that detunes electronic and photonic excitations and can be interpreted to some extent as a collective renormalization of the electronic excitation or vice versa the photonic energy.
Hereby $ \textbf{r}_{01}^{\{\textbf{R}_n\}} $ depends parametrically on the nuclear configuration in terms of one \textit{individual} molecular state but identical for all molecules. Assuming the molecular system consists of different species or molecular configurations, one can partially symmetrize those to recover a collective behavior for each set of identical configurations~\cite{galego2015,feist2017polaritonic}. The obtained energy splitting fits the lowest order contribution, that is $n=0$, of the Tavis-Cummings model~\cite{garraway2011} up to the factor $(\omega-\delta)/\omega$ with 
\begin{align}\label{taviscummings}
\Delta E_{n=0}^{TC} = \sqrt{ \delta^2 + (\sqrt{N_e}\sqrt{2\omega}\lambda)^2 }.
\end{align}
It becomes evident that within such a strongly simplified expansion, that is a single photonic and electronic excitation, the resulting solution will vary depending on the selected coupling form, that is, whether we assume the length or momentum picture. The only exception occurs exactly on resonance or for a completely converged expansion (see Sec.~\ref{numchecks}).
In general, the coefficients $c^{i=1,\pm}_{01}(\tilde{\delta}(\{\textbf{R}_n\}),\tilde{\Omega}(\{\textbf{R}_n\}))$ depend on the number of particles, not only via the Rabi-splitting but also due to the collective detuning. Furthermore, their behavior is essential close to conical intersections as they determine the non-adiabatic nuclear-polariton coupling elements $\langle \psi^l_{\nu} \vert \nabla_{\textbf{R}_j}\vert \psi^k_{\mu} \rangle_e$ discussed in Sec.~\ref{roothaanhall} (see  Eq.~\eqref{nonadiabatic_elements_nuclpol}).
 Such a detuning as shown in Eq.~\eqref{lambshift} caused by the collectivity of a frequency-shift was measured \cite{rohlsberger2010collective,garcia2015light} and extensively discussed in Ref.~\cite{friedberg1973frequency}. As a consequence of the collective detuning, the resonance does no longer coincide with zero detuning $\delta$ but appears for values  $ \delta  = \omega - \Delta E_e = -\mathcal{L}$.
 As a consequence of the quadratic coupling, the collective molecular participation or the coupling itself has to be significant otherwise the effect is negligible. However, within strong coupling, as for example $\omega_\alpha \sim 0.1$, $\sqrt{N_e}\lambda \sim \sqrt{100}\cdot0.01 \sim  0.1$ we can see that $\mathcal{L}$ is on the order of $\tilde{\Omega}$ and consequently non-negligible.\\
 In combination with the remaining block involving the counter-rotating contributions
 \begin{align*}
 \left(
 \begin{matrix}
 H_{BO,00}^{l=0} - \frac{N_e}{2} \Delta^{00} - E & \nabla_{e}^{01} \cdot \nabla^{10} \\
 \nabla_{e}^{10} \cdot\nabla^{01} & H_{BO,11}^{l=1} - \frac{N_e}{2} \Delta^{11} - E
 \end{matrix}
 \right)
 \cdot
 \left(
 \begin{matrix}
 c_{00}\\
 c_{11}
 \end{matrix}
 \right)
~,
 \end{align*}
 the collective many-body (cavity-matter) energetic levels are given by
 \begin{align}
 E_0' &= \bar{E} - \frac{1}{2} \sqrt{\tilde{\Delta}^2 + \tilde{\Omega}^2 } \label{eigen1}\\
 E_1^{-} &= \bar{E} - \frac{1}{2} \sqrt{\tilde{\delta}^2 + \tilde{\Omega}^2 } \label{eigen2}\\
 E_1^{+} &= \bar{E} + \frac{1}{2} \sqrt{\tilde{\delta}^2 + \tilde{\Omega}^2 } \notag\\
 E_2' &= \bar{E} + \frac{1}{2} \sqrt{\tilde{\Delta}^2 + \tilde{\Omega}^2 }\notag
 \end{align} 
 where 
 \begin{align*}
 \bar{E} &= E_0 + \frac{\Delta E_e + \omega}{2} + \mathcal{L}\\
 \tilde{\Delta} &= \Delta E_e + \omega + \mathcal{L}~.
 \end{align*}
Here,  $E_1^{-}$ and $E_1^{+}$ correspond to lower and upper polariton with conserving excitations, that is, the states are superpositions of excited matter or cavity but never both. In contrast, $E_0'$ describes the renormalized ground-state of the correlated system while $E_2'$ is connected to the coupled state where electronic and photonic system are excited simultaneously. If we drop the excitation non-conserving couplings, that is, the coupling from no excitation at all to both subsystems are excited $\nabla_{e}^{01} \cdot \nabla^{10}$ and $\nabla_{e}^{10} \cdot\nabla^{01}$, $E_0'$ and $E_2'$ do not change beside energetic shifts via $\mathcal{L}$.
The single photon polariton does naturally include counter-rotating terms within this single excitation subspace as presented above. Whether the single photon polariton approximation is reasonable depends on the relevance of multi-photon processes. As long as we expect them to be negligible the single photon polariton approximation is expected to capture the essential physics. This gives a very intuitive criterion to judge the quality of this minimal model. But, as will be discussed in Sec.~\ref{numchecks}, this intuition can sometimes be a little misleading as the quantitative agreement depends on the interplay of photonic and electronic excitations as well as the observables of choice.\\

\noindent
Next we consider what the single photon polariton can teach us about the alleged Rabi-phase transitions in coupled matter-photon systems.
From Eqs.~\eqref{eigen1} and \eqref{eigen2}, we can observe
  \begin{align*}
  E_1^{-} - E_0' &> 0,~\forall \vert\tilde{\Omega}\vert<\infty~.
  \end{align*}
That is, what was the ground state at $\lambda =0$ is connected to the ground state for all $\lambda >0$. Only in the thermodynamic limit of $N_e \rightarrow \infty,~\vert\tilde{\Omega}\vert \rightarrow \infty $, the ground-state becomes degenerate with the lower polariton. Performing the rotating-wave approximation and solving the following Tavis-Cummings model, namely shifting the first polariton by Eq.~\eqref{taviscummings} will result in a crossing of lower polariton and the non-sensitive ground-state, that is, $ E_1^{-}-E_0 < 0 ~\forall \lambda > \lambda_c $ with a critical coupling $\lambda_c$. This type of a Rabi phase transition is not observed within the single photon polariton. In contrast to the out-of-equilibrium super-radiance transition which has been experimentally verified, a phase-transition in equilibrium is still highly debated~\cite{dicke1954coherence,knight1978,viehmann2011superradiant,griesser2016depolarization,de2018cavity}. Also note that neglecting the counter-rotating contributions is only valid for $\tilde{\Omega} \ll \tilde{\Delta}$.
Although the single-photon collectivity that is captured by the single photon polariton approximation is dominant, it only corresponds to the linear contribution for collective effects.
Nonlinearities can be introduced by matter-photon interaction where also the self-polarization can be substantial and/or multi-photon participation which can lead indeed to drastic transitions as will be presented in Sec.~\ref{numchecks} and has been observed also in Refs.~\cite{flick2015,flick2017ab}. Multi-photon participations are not represented in the above single-photon approach and a non-trivial collectivity leading to a ground-state transition can consequently not be ruled out from the above simplifications. As for example presented in Refs.~\cite{knight1978,viehmann2011superradiant,griesser2016depolarization,de2018cavity}, the phase-transition does drastically depend on the interplay between self-polarization, bilinear coupling and Coulomb interaction. The following section will show that these different factors call into doubt the reliability of few-level approximations. An extended discussion of this question including an accurate investigation of the electron-photon coupled system by high-level renormalization-group techniques is currently in progress~\cite{cs2018superradiance}.

\section{Numerical Details}\label{numchecks}
\noindent
Finally, after we have simplified the original nucleus-electron-photon problem by the explicit and then the single photon polariton approximation, we want to investigate the reliability of these approximations. We will do so by considering a real-space system that we then approximate by a different number of basis functions for the photonic and electronic subspaces. To keep the numerical calculation tractable we will focus on the polaritonic subsystem and freeze the nuclear coordinates. By assumption, it is only the electronic subsystem that is directly affected by the coupling to the photons and the nuclei merely feel the presence of the photonic modes as a change in the potential-energy surfaces. We therefore investigate the accuracy of the different levels of approximation in terms of the polaritonic observables and wavefunctions only. Before we do so, we want to give a brief recapitulation of the general explicit polariton approximation and its numerical implementation. This work-flow can be straightforwardly extended  beyond the explicit polariton to the full Born-Huang expansion of Eq.~\eqref{eq:electroniccompact} and \eqref{eq:nuclearcompact}.

\subsection{Computational scheme for the explicit polariton}\label{compschemetext}
\begin{figure}
	\includegraphics[width=0.9\linewidth]{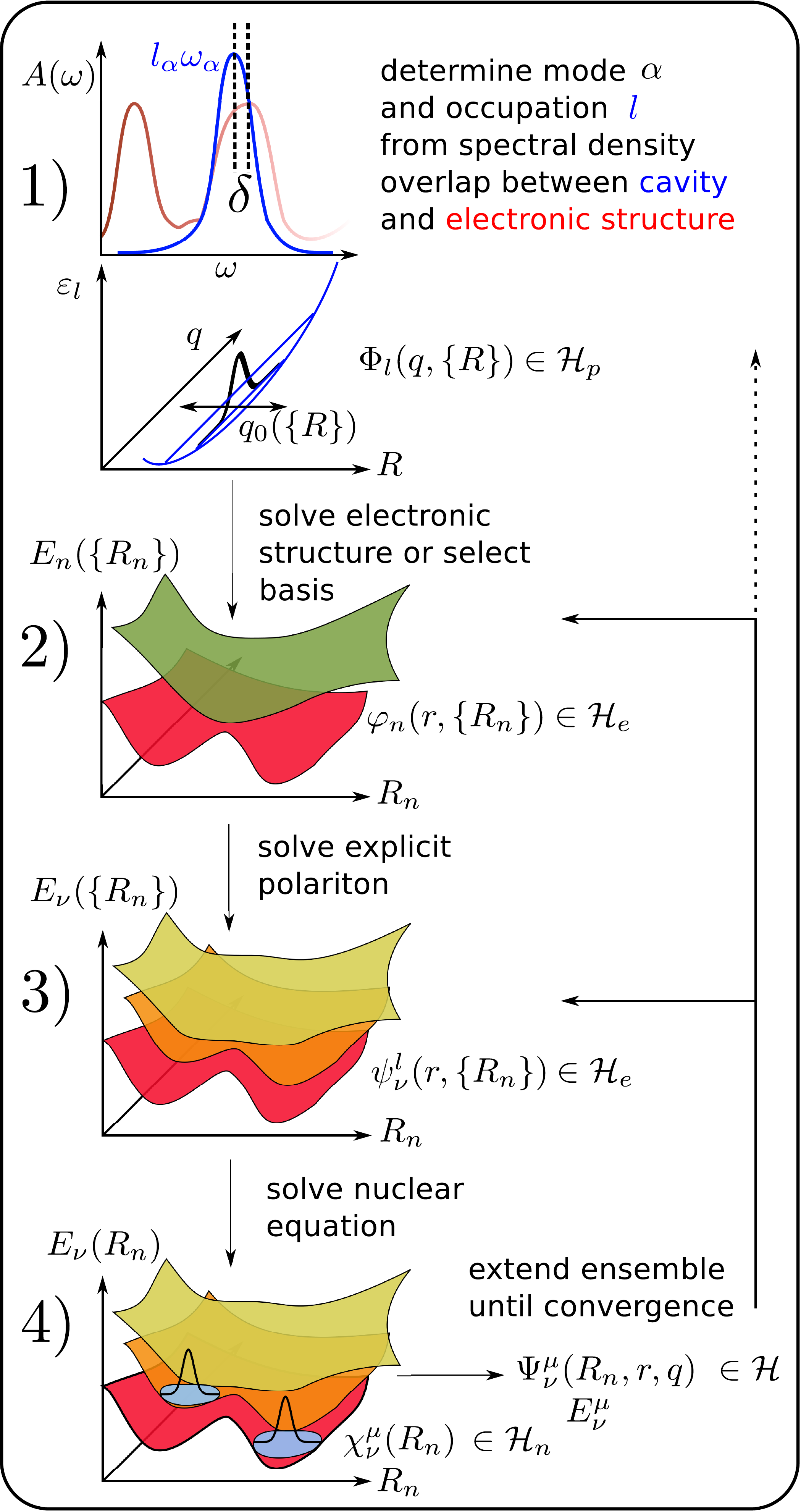}
	\caption{Computational scheme for the explicit polariton solution as discussed in Sec.~\ref{compschemetext}. Here $\mathcal{H}_{p}$ is the photonic subspace, $\mathcal{H}_{s}$ is the electronic subspace and $\mathcal{H}_{n}$ is the nuclear subspace. Step 1) is a purely conceptional selection of energetic domains, 2) corresponds to the very common process of solving the electronic many-body problem with parametric nuclear dependence while 3) represents the essential and computationally novel step. Solving the polaritonic step 3) can for example be efficiently done with slight adjustments of available Floquet implementations. Step 4) corresponds to the familiar nuclear solution with potentially novel and adjusted non-adiabatic couplings as discussed.}
	\label{flow}
\end{figure}

\noindent
The derived equations incorporate the photonic degress of freedom for equilibrium QED calculations in a consistent way into the common Born-Oppenheimer quantum chemistry approximation. The additional computational effort remains limited as the photonic subspace can be represented as bare excitations of electronic $\varphi_n(\textbf{r},\{\textbf{R}_n\})$ and nuclear $\chi_\nu^\mu(\textbf{R}_n) $ wavefunctions. Within, for example, configuration-interaction or coupled-cluster techniques~\cite{purvis1982full}, the excitations are part of the minimization procedure of the fermionic equation anyway and can be used directly within the minimization procedure as schematically presented in Fig.~\ref{flow}. The computational work flow of the explicit polariton approximation then becomes:\\
In the first instance $1)$, we have to clarify how many modes with corresponding frequency $\omega_\alpha$ are essential to describe the field-matter interaction. Commonly a single mode is assumed but there are situations where more are relevant~\cite{george2016}. This can be approached by, for example, comparison of cavity or nanostructure and matter spectral functions. It determines which modes and frequencies are essential as well as the nature of matter excitations, that is, dominantly nuclear or electronic excitations.\\
In instance $2)$, we use the insight from step $1)$ to calculate or select the necessary set of electronic states for a sufficient domain of parametric nuclear positions $\{\textbf{R}_n\}$. The selected set of electronic states does not have to be a set of eigenstates of the electronic problem but it will render the following step simpler.\\
Next, in step $3)$, those bare electronic states are mixed by Eqs.~\eqref{maineq} or \eqref{rothaan_hall} into coupled electron-photon states (polaritons). As discussed before, due to the structural similarity to Floquet theory, small adjustments in existing implementations can lead to efficient solutions of this step with marginal additional effort.\\
Finally, we solve the nuclear component \eqref{nuclBO} in step $4)$ on the polaritonic energy-surfaces, potentially including non-adiabatic coupling elements from electronic, photonic or mixed excitations (see the full Born-Huang expansion Eq.~\eqref{nsub}).\\
This procedure has to be iterated until self-consistency is reached. Higher excitations are typically stronger delocalized which is more likely to appear with photonic interaction. The ensemble of parametric nuclear values $\{\textbf{R}_n\}$ in the polaritonic subspace has to be potentially adjusted. Beyond the long-wavelength approximation, this would also take place in the photonic subspace.\\[1\baselineskip]
\noindent
Alternatively for step $3)$, the Coupled-Cluster technique or any other modern quantum-chemistry scheme to deal with electron correlations could be directly applied to a polaritonic Slater determinant in Born Oppenheimer approximation $ \tilde{\Psi}_0 (\{\textbf{R}_n\}, \textbf{r}, \textbf{q}) \approx \psi^0_{0} (\textbf{r},\{\textbf{R}_n\}) \Phi_0(\textbf{q},\{\textbf{R}\}) $ with a possible cluster excitation operator for electronic and photonic system
\begin{align*}
\hat{T}_{c} &= \exp \left[\sum_n\hat{T}^{(n)} + \sum_m \hat{P}^{(m)} \right]\\
\hat{P}^{(m)} &= \sum_{\alpha_1,...,\alpha_m} c_{\alpha_1,...,\alpha_m} \hat{a}^\dagger_{\alpha_1} ...  \hat{a}^\dagger_{\alpha_m}~.
\end{align*}
\noindent
Recently, together with others, we presented how an efficient implementation of the QED ground-state correction can be accomplished by means of quantum-electrodynamical density-functional theory within the optimized effective potential (OEP) approach~\cite{flick2017ab}. There, the Kohn-Sham system is factorizing photonic and electronic system, which accounts for the classical shift of the Harmonic oscillator basis. The remaining quantum nature of the photonic orbitals is contributed by an effective potential which can be derived from perturbative corrections $\Phi^{(1)}_{i\sigma,\alpha}$ to the Kohn-Sham orbitals $\phi_{i\sigma}$~\cite{flick2017ab}. Those corrections carry a single photon and are the solution to a Sternheimer response equation, that is, they incorporate the effect of the response of the system to photonic fluctuations. We can identify a structural similarity of polaritonic associated states $\psi^l_{\nu}(\textbf{r},\{\textbf{R}_n\})$ and perturbative orbital corrections. This is especially prominent in the mode occupation of section \ref{phtonic_obs}.

\subsection{Numerical benchmarks}\label{numbench}
\noindent
Since the accuracy of the nuclear Born-Oppenheimer approximation is well known, we focus here on the polaritonic contribution $3)$ to gather additional physical insight for the presented approach, that is, we skip the final step in Fig.~\ref{flow}.
Additionally, $3)$ is the step that connects first-principles quantum-chemical to quantum-optical model systems and we will see how a reduced set of electronic states affects the qualitative and quantitative behavior.\\
An efficient implementation of the explicit polariton can start from Eq.~\eqref{rothaan_hall}. To allow for an exact reference, that is, distill the effect of photonic interaction, we solve the electronic system by exact diagonalization and use the unperturbed states in Eq.~\eqref{rothaan_hall}. 
The coefficient-matrices $c^\nu_{nl}$ relate a set of bare electronic states $\varphi_n(\textbf{r})$ with $n\in \{0,1,...,n_{max}\}$ to mode excitations $l\in \{0,1,...,l_{max}\}$. The full space resolved polaritonic eigenfunctions $\psi_\nu (\textbf{r}) = \{ \psi_\nu^0(\textbf{r}), \psi_\nu^1(\textbf{r}), \psi_\nu^2(\textbf{r}), ..., \psi_\nu^{l_{max}}(\textbf{r}) \}$ are represented in the limited set of bare electronic eigenfunctions $\varphi_n(\textbf{r})$.
The resulting eigenstates represent the electronic as well as photonic contribution for each polariton eigenstate $\nu$. For example, the ground-state density of the correlated electron-photon system is given by
\begin{align*}
n_{\nu = 0}^{\lambda>0}(\textbf{r}) = &\sum_{n,n'=0}^{n_{max}} \int d\textbf{r}_2 d\textbf{r}_3 ... d\textbf{r}_{N_e}\varphi^*_{n}(\textbf{r},\{\textbf{R}_n\})\notag \\
&\varphi_{n'}(\textbf{r},\{\textbf{R}_n\}) \sum_{l=0}^{l_{max}} c^{\nu =0}_{ln}c^{\nu =0}_{n'l}~.
\end{align*}
The density is an important observable in quantum chemistry that allows for intuitive interpretations and gives information, for instance, on the nature of chemical bonding~\cite{szabo1989}. The change of the real-space resolved density in the correlated matter-photon system can be helpful in interpreting the effect of the emergence of polaritonic states. Such changes do also directly affect the non-adiabatic coupling elements and can prove useful to get a more detailed understanding of the influence of cavity photons on specific chemical processes, as for example transition states, solvation energies, and involved processes within energy transfer. A further observable of quantum-chemical interest is the energy-difference between ground and first excited state $\Delta E_{0\rightarrow1}$ as those energies present the surfaces on which the nuclear wavepackets move dominantly.
On the other hand, of quantum-optical interest is for example the energy-difference between first and third excited correlated state $\Delta E_{1\rightarrow3}$ which represents the Rabi-splitting in our numerical example for weak coupling of Sec.~\ref{2dgaas}. The cavity mode occupation $ \langle \hat{a}^\dagger \hat{a} \rangle $ as presented in Sec.~\ref{phtonic_obs} is an additional observable of interest. It is important to note that except of the density all the shown observables are integrated quantities. It is usually much more involved to accurately capture non-integrated (space-resolved) quantities than integrated quantities. This will become clear in the numerical example discussed in Sec.~\ref{2dgaas}. So the accuracy of the approximation strongly depends on the quantity that is considered and simple models, while reliable for certain quantities, are less so for other observables.\\
For computational convenience, we map the coefficient matrix on a vector
\begin{align*}
c^\nu_{nl} \rightarrow c^\nu_{m}~, ~m = n\cdot l_{max} + l
\end{align*}
to recast the explicit polariton into a diagonalizable matrix equation.

\subsection{2D GaAs quantum ring}\label{2dgaas}
\noindent
Now we will investigate in detail the different possible levels of approximations. By this we mean that we consider the numerical convergence with respect to the electronic $n\in \{0,1,...,n_{max}\}$ and the photonic sub-space $l\in \{0,1,...,l_{max}\}$.
The computational demand in a truncated subspace is significantly lower than an exact diagonalization in the full electron-photon space while reaching a high level of accuracy up to strong coupling. This is a consequence of the fact, that the electronic eigenstates are a well suited basis-set for equation \eqref{rothaan_hall}. It changes if the electron-photon correlation extends into the ultra-strong coupling where the electronic structure is drastically distorted by the correlation. The limited sub-set $n^{\lambda \ll 1}_{max}$ is then no longer sufficient to describe those distortions and we have to extend our minimal set $n^{\lambda \ll 1}_{max} < n^{\lambda \sim 1}_{max}$ to describe the correlated system. This will clarify that a few-level approximation is problematic by construction in the (ultra-)strong coupling domain.
We investigate a model of a GaAs quantum ring introduced in previous works~\cite{flick2015,flick2017b,flick2017ab}.  Hereby, one electron is trapped in a two-dimensional ``Mexican hat'' potential
\begin{align*}
\hat{H}_{en} = \frac{\xi_1}{2} (\hat{\textbf{r}}\cdot\hat{\textbf{r}}) + \xi_2\cdot\exp\left[-\frac{(\hat{\textbf{r}}\cdot\hat{\textbf{r}})}{\xi_3^2}\right]
\end{align*}
with $\xi_1 = 0.7827,~ \xi_2 = 17.70$ and $ \xi_3 = 0.997$.
We couple this system to a single mode in resonance to the first excitation $\omega_\alpha = E_1^{\lambda=0}-E_0^{\lambda=0}$ with polarization in the diagonal direction $\boldsymbol\lambda = \lambda (\textbf{e}_x + \textbf{e}_y)$. We then vary the effective coupling strength. The electronic structure is solved on a 201-by-201 grid with spacing $\Delta x = 0.1$ and derivatives approximated by forth-order finite-differences. The photonic excitation space is incorporated by 40 Fock number-states for weak and 80 for ultra-strong coupling. For additional information, we refer the reader to Ref.~\cite{flick2017ab}.

\subsubsection{Weak coupling solution}
\begin{figure*}
	\includegraphics[width=1\textwidth]{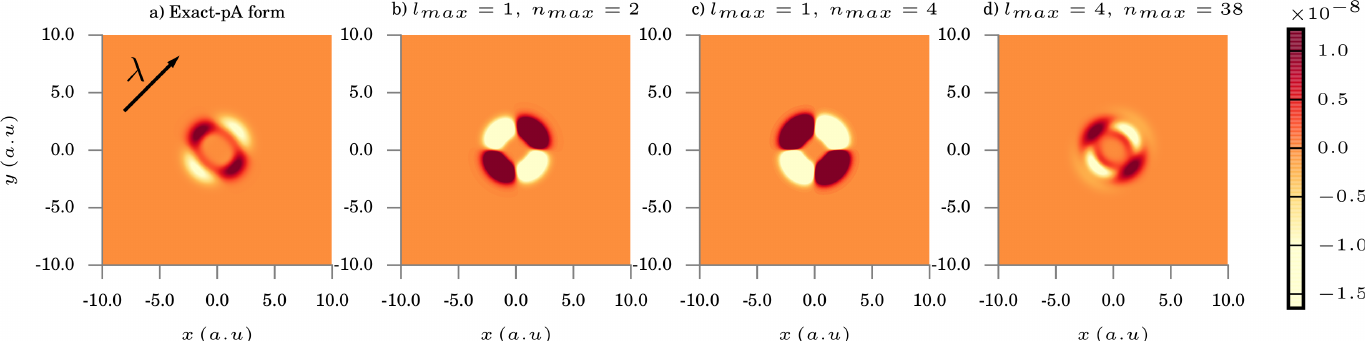}
	\caption{Weak coupling. Photonic influence on the ground-state density $\Delta n(\textbf{r}) = n_{\lambda>0}(\textbf{r}) - n_{\lambda=0}(\textbf{r})$. Exact solution a) with a 40 photon Fock-space. The two-level approximation is not capable to qualitatively reproduce the correct physical behavior, even for weak couplings.}
	\label{gaas1}
\end{figure*}

\noindent
The exact solution has two limiting cases with respect to the effective coupling. For a weak to strong coupling $\lambda = 0.005$, the system remains dominantly in its initial configuration and avoids the self-polarization potential, therefore the density is elongated perpendicular to the polarization. This can be illustratively seen in the exact solution Fig.~\ref{gaas1} a). While the linear component of the photonic interaction $\omega\hat{q}\boldsymbol\lambda\cdot \hat{\textbf{R}}$ favors density-accumulation along the polarization, the quadratic self-polarization potential $(\boldsymbol\lambda\cdot \hat{\textbf{R}})^2$ diminishes it. For the ground-state, where the bilinear contribution becomes relevant in second order perturbation theory, the correlated wavefunction is calculated in a potential with two competing contributions. Their relation is sensible to the frequency and can consequentially be adjusted accordingly. In this weak-coupling example, the quadratic part is dominating and is essential to describe the qualitative behavior of the correlated ground-state. Therefore ignoring the self-polarization (as done by construction in simple two-level approximations) would lead to a qualitatively wrong result.
The photonic contribution is here a weak perturbation of the system and can be very accurately recovered by effective single excitation approximations, for example the exact-exchange optimized-effective potential approach in the context of quantum-electrodynamical density-functional theory presented in Ref.~\cite{flick2017ab}.\\[1\baselineskip]
The polariton equation effectively mixes momenta of different electronic eigenstates. As a consequence of the radial symmetry, there are two degenerate first excited solutions, one with dominant momentum parallel and one with dominant momentum perpendicular to the field polarization. The minimal basis set for electronic excitations has to include both states, otherwise the dominant contribution is not properly captured. If we take this into account, therefore extend the electronic excitations to two or more, already the single photon polariton $l_{max}=1$ does recover the exact energetic structure for weak coupling quite accurately as presented in Tab.~\ref{table2}.\\[1\baselineskip]
\begin{table}
	\label{table2}
	\begin{tabularx}{\columnwidth}{l|l|l|X|X|X}
		$l_{max}$ & $n_{max}$ & $\lambda$ & $\Delta E_{0\rightarrow 1}$ & $\Delta E_{1\rightarrow3}$ & mode occupation\\[0.1\baselineskip]  
		\hline 1 & 2 & 0.005 & $0.1224009$ & $0.0054393$ & $0.0001180$ \\ 
		1 & 8 & 0.005 & $0.1223674$ & $0.0054417$ & $0.0001180$ \\ 
		1 & 19 & 0.005 & $0.1223514$ & $0.0054419$ & $0.0001334$ \\[0.4\baselineskip] 
		2 & 2 & 0.005 & $0.1223748$ & $0.0054324$ & $0.0001182$ \\ 
		2 & 4 & 0.005 & $0.1223403$ & $0.0054369$ & $0.0001183$ \\
		2 & 8 & 0.005 & $0.1223403$ & $0.0054369$ & $0.0001183$ \\ 
		2 & 19 & 0.005 & $0.1222896$ & $0.0054356$ & $0.0001336$ \\[0.4\baselineskip]  
		4 & 38 & 0.005 & $0.1222861$ & $0.0054355$ & $0.0001343$\\[0.4\baselineskip]  
		ex-pA & - & 0.005 & $0.1222855$ & $0.0054355$ & $0.0001346$ \\[0.1\baselineskip] 
		ex-dE & - & 0.005 & $0.1222855$ & $0.0054355$ & $0.0001183$ \\[0.1\baselineskip] 
		\hline 1 & 2 & 0.4 & $0.1193664$ & $1.3520221$ & $0.0190439$ \\[0.4\baselineskip]  
		4 & 8 & 0.4 & $0.1910540$ & $0.0610576$ & $0.3205813$ \\  
		4 & 38 & 0.4 & $0.0116341$ & $0.1332598$ & $0.2719931$ \\[0.4\baselineskip]   
		19 & 38 & 0.4 & $0.0059610$ &$0.1084440$ & $0.4976897$ \\[0.4\baselineskip]
		ex-pA & - & 0.4 & $0.0020865$ &$0.0992033$ & $0.4571209$\\[0.1\baselineskip] 
		ex-dE & - & 0.4 & $0.0020814$ &$0.0990272$ & $3.1917314$
	\end{tabularx} 
	\caption{Energetic convergence of the explicit polariton equation \eqref{rothaan_hall} in relation to the exact solution for different photonic $l_{max}$ and electronic $n_{max}$ sets of excitations expressed in atomic units. The exact results (ex-(pA/dE)) are calculated using a number of 40 (weak coupling) or 80 (ultra-strong coupling) photonic Fock number-states. The different mode occupations emerge from the coupling-form dependent interpretation of the modes (see Sec.~\ref{phtonic_obs}) in length (ex-dE) or momentum (ex-pA) form. For a detailed analysis and interpretation, we refer the reader to the text.}
\end{table}
Considering the first three rows in Tab.~\ref{table2} which represent the single photon polariton approximation, we observe that increasing the number of electronic states does not drastically improve the accuracy at first glance. While the chemically important ground-excited energy-difference does slightly improve, the polariton split and the mode occupation do not or even become worse. If we allow for at most double photonic occupation, this picture changes. Increasing the electronic set does then indeed improve the energies. We can intuitively understand this from the following argument. As long as we allow just for a single mode excitation, the probability of reaching higher excited states is vanishingly small. Including them nevertheless can then lead to inaccuracies since their effect is not properly captured in the approximation. The moment we allow our theoretical description to have two or more mode excitations, we are able to reach higher excited states and those states can now participate by contributing energy and momentum to the correlated solution.\\[1\baselineskip]
Especially interesting is that the mode occupation introduced in Sec.~\ref{phtonic_obs} accurately reproduces the one in length form (ex-dE) as long as the the set of eigenstates include just the first set (1S, 1P, ...). As we incorporate the next order (2S, 2P, ...) $n_{max}>8$, the mode occupation resembles the one in momentum form (ex-pA). While other integrated quantities change marginally, the mode occupation does not only drastically depend on the number of excited states, it even qualitatively changes its physical interpretation depending on the selected electronic set.\\

\noindent
If we compare in Fig.~\ref{gaas1} the electronic density influence of the smallest possible set, that is, an effective two-level system with $n_{max}=2$ shown in b), to the exact solution a), we find a qualitatively different behavior. The single photon polariton approximation with $n_{max}=2$ cannot even capture qualitatively the right physical behavior even for weak couplings. Although the single photon polariton does only couple by bilinear components $\nabla^{lk} \sim \hat{p}$, it partially incorporates the self-polarization by construction. As discussed before in Sec.~\ref{relselfpol}, an effective two-level system is not able to properly capture the self-polarization and the same holds true in momentum form. However, if we stay within the single photon polariton $l_{max}=1$ but extend the electronic set to ground-state plus four excited levels $n_{max}=4$, the solution presented in c) does switch into the correct orientation. This puts in evidence once more that self-polarization and $\textbf{A}^2$ part cannot be interchanged or regarded as similar, especially in a truncated electronic set. By further increasing the number of mode and electronic excitations, we start to quantitatively recover the exact result as can be seen in Fig.~\ref{gaas1} d). As a result, it becomes evident that a strongly simplified electronic structure is not sufficient to capture the physical behavior of the correlated ground-state. That the electronic set has to be extended quite a bit is also caused by the distorted elliptic shape of the exact solution which is demanding to recover from a superposition of angular momentum states. Additionally, the set of states should be chosen consistently with respect to the distribution of momenta. A selection which is slightly larger but not balanced is in our experience not beneficial.\\

\noindent
As discussed in Sec.~\ref{pzwequiv}, an alternative approach to address changes in the ground state is to parametrically incorporate parts of the interaction in the electronic equation. In this way we avoid the expansion in bare electronic states by adjusting the electronic equation and by introducing further parametric dimensions $\{q_\alpha \}$. Irrespective of whether one chooses this cavity Born-Oppenheimer approach~\cite{flick2017b} or employs the here employed expansion in bare electronic eigenstates, the common assumption of an electronic two-level system is not sufficient to capture the observed effects.

\begin{figure}
	\includegraphics[width=1\columnwidth]{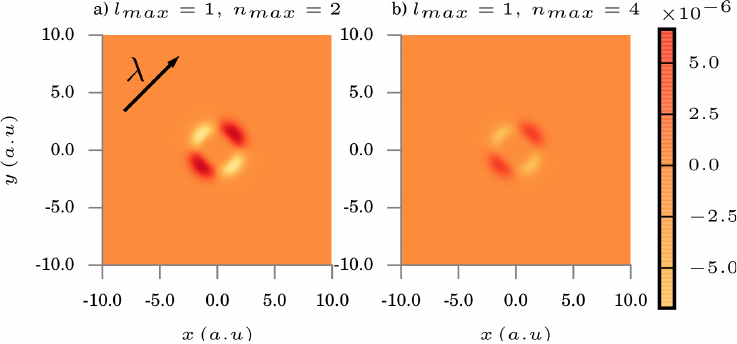}
	\caption{Weak coupling. Convergence of the lower polariton density $\Delta n_{P^-}(\textbf{r}) = n_{P^-}(\textbf{r}) - n^{\text{ref}}_{P^-}(\textbf{r})$ with respect to the reference solution with $l_{max}=4,~n_{max}=38$. The two-level approximation a) can as expected qualitatively reproduce the results while we observe deviations on the next perturbative order. Increasing the number of electronic states b) improves the quantitative results. Notice that the counter-rotating component is included.}
	\label{gaas2}
\end{figure}

\noindent
The density of the lower polariton can be captured to a sufficient amount by the two-level approximation as presented in Fig.~\ref{gaas2}. The dominant contribution consists of superposition of ground and first excited state and higher order corrections such as secondary radiative interactions, namely the interaction between 1P and 1D, correct the two-level approximation.\\
The moment we are interested in ground-state observables such as the electronic density, essential for chemical reactions, the electronic two-level approximation is not sufficient to recover qualitatively the correct behavior. Beyond this, depending on the observable of interest and coupling, typical approximations can be sufficient but have to be questioned on a case by case basis. Strongly simplified models become problematic as the electronic structure becomes distorted by the photonic environment. Those effects are enhanced in the collective ensemble according to appendix \ref{colcoupling}.

\subsubsection{Ultra-strong coupling transition}
\begin{figure*}
	\includegraphics[width=1\textwidth]{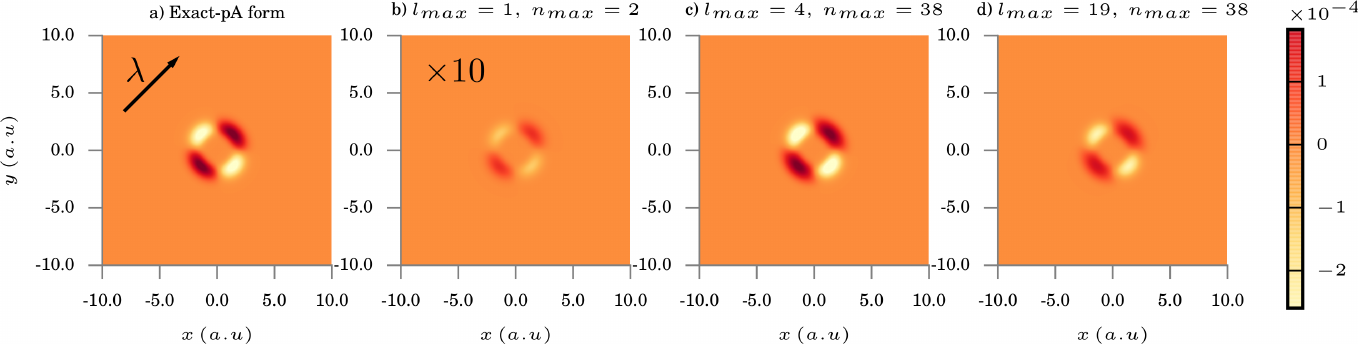}
	\caption{Ultra-trong coupling. Photonic influence on the ground-state density $\Delta n(\textbf{r}) = n_{\lambda>0}(\textbf{r}) - n_{\lambda=0}(\textbf{r})$. Exact solution a) with a 80 photon Fock-space. The effective two-level approximation b) $l_{max}=1,~n_{max}=2$ is amplified by a factor $10$ to compare with other solutions.}
	\label{gaas3}
\end{figure*}

\noindent
For ultra-strong coupling $\lambda = 0.4$ and therefore $\lambda/\sqrt{2\omega} = 0.8$, the photonic contribution is no longer a weak perturbation but does drastically reshape the electronic landscape. This is very illustrative in the effect on the electronic density as presented in the exact solution Fig.~\ref{gaas3} a). Here, $\Delta n(\textbf{r})$ does no longer avoid the direction of polarization but is distorted along the polarization axis. The bilinear contribution is therefore dominating over the self-polarization and we enter a regime which we could assign to a super/sub-radiant phase as briefly discussed in Sec.~\ref{modelsystem}.\\
Although the explicit polariton in its perturbative character becomes problematic, including a sufficient number of excited states, which is still negligible in relation to the full electronic Hilbert-space of 40401 eigenstates, can capture the ultra-strong coupling twist. Nevertheless, in this limit it becomes extensively cumbersome to recover the correct energies and densities, therefore the single photon polariton is no longer a satisfying approximation even for the integrated quantities.\\
The effective two-level approximation b) is essentially scaling the weak coupling solution, deviating strongly in the integrated observables. The energies and mode-occupations in Tab.~\ref{table2} deviate drastically from the exact solution. Recovering the correct energies is now demanding even for larger electronic sets and convergence is slow. Especially, the qualitative behavior of the density demands the next manifold of excited states, that is, 2S, 2P and so on, therefore $n_{max} \gg 8$. Furthermore, subset c) and d) in Fig.~\ref{gaas3} elucidate that it can be even counter-beneficial for the density while beneficial for integrated quantities to incorporate a large amount of photonic excitations without increasing the electronic set. From Tab.~\ref{table2} and Fig.~\ref{gaas3}, we conclude that within the ultra-strong coupling domain, a fully consistent method is demanded that is able to incorporate a large part of the electronic and photonic Hilbert space.\\

\section{Summary and Conclusion}\label{summary}
\noindent
We have presented a consistent and comprehensive derivation of approximation strategies for coupled nucleus-electron-photon systems based on the Born-Huang expansion. We highlighted the connection between length and momentum form with the help of a shifted harmonic oscillator basis for the photonic subsystem, provided an alternative perspective on the relevance of the self-polarization contribution, discussed connections to Floquet theory and deduced an approach how to solve the fully correlated nucleus-electron-photon many-body system for finite and periodic systems. Especially structural similarities to Floquet theory imply the possibility of efficient implementations with marginal effort and furthermore ways to circumvent common problems of Floquet theory. The resulting generalization of the Born-Oppenheimer approximation to include photons we have named the explicit polariton approximation. Since we could solve the photonic subsystem analytically the numerical costs of this approximation is comparable to common Born-Oppenheimer calculations. If we restrict the number of photonic excitations to one, the resulting simplified equations we then called the single photon polariton, which we were able to solve analytically when also the electronic excitations were restricted. This allowed us to connect to well-known analytical models of quantum optics and discuss implications, for example, for superradiant phase transitions. Finally we investigated the reliability of the explicit polariton approximation for weak to strong and ultra-strong coupling and for different numbers of basis functions. We found that for integrated quantities such as excitation energies for weak couplings already the simple quantum-optical models were very accurate but mode occupation and spatially resolved quantities such as the density needed more basis functions to be at least qualitatively correct depending on the state of choice. On the other hand, increasing the number of basis functions for the electronic subsystem while keeping the photonic basis limited to one excitation led in turn to worse results in certain integrated quantities. A balanced extension of basis functions in electronic and photonic subspace is necessary. In the ultra-strong coupling regime the single photon polariton approximation, also for the integrated quantities, was qualitatively wrong and a consistently expanded basis is essential.\\

\noindent
The presented results clearly highlight how first-principle approaches based on non-relativstic QED in the long-wavelength limit and models of quantum optics and polaritonic chemistry are connected. They show how simple models can be extended to converge to the first-principle results by including in a consistent manner more and more basis functions going beyond few-level approximations. To go beyond simplified models or at least to be aware of the limitations of these models becomes increasingly important in the context of polaritonic chemistry and material sciences~\cite{thomas2016,galego2015,flick2017,flick2017ab,ribeiro2018polariton}. A theoretical description of strong matter-photon coupling to influence material properties clearly needs a description of the matter subsystem that can genuinely represent the physical and chemical processes and a simplified few-level description might not be enough to capture the complex processes that can happen in real systems. The presented explicit polariton approximation together with similar generalizations of quantum-chemical methods~\cite{flick2017b, galego2015} as well as generalizations of many-body methods such as density-functional theory for coupled matter-photon systems~\cite{ruggenthaler2014, ruggenthaler2015, flick2017ab} paves the way for such studies from first principles~\footnote{Implementations along the lines of modified Floquet and quantum-dynamical density-functional theory will be implemented in the highly parallelized Octopus code}. Furthermore, the presented unifying framework highlights the similarities between different settings of strong light-matter interactions, for example, due to plasmonic nanostructures or due to external driving, and even between different fields of quantum physics, such as quantum chemistry, quantum optics and quantum topological matter out of equilibrium. It therefore presents a platform to exchange ideas and concepts between different areas of research and can help in guiding the development of novel quantum technologies.

\begin{acknowledgments}
\noindent
We would like to thank Heiko Appel, Arunangshu Debnath and Johannes Flick for insightful discussions and acknowledge financial support from the European Research Council (ERC-2015-AdG-694097), and the European Union's H2020 program under GA no.676580 (NOMAD).
\end{acknowledgments}

\appendix

\section{Effective coupling strength}\label{colcoupling}
\noindent
While the fundamental coupling strength as defined in Eq.~\eqref{fundamentalcoupling} is determined by the physical situations, for example, the properties of the cavity or the surrounding environment, one can still define an effective coupling strength that takes into account collective effects. The simplest example to rationalize this is the Tavis-Cummings model~\cite{garraway2011},  of many two-level systems as discussed in Sec.~\ref{modelsystem}. Here the rotating-wave approximation allows to calculate the Rabi splitting (which is used as a measure of how strong the coupling is) by
\begin{align}
 \Omega_{\rm R} \sim |\langle g| \hat{\textbf{r}}| e \rangle| \sqrt{\frac{2 N 
\omega}{V}}  \sqrt{n_{ph}+1}, \nonumber
\end{align}
where, as discussed the cavity quantities (the effective volume $V$ and the frequency $\omega$), the dipole-transition element between the ground $| g\rangle$ and the excited state $| e \rangle$ of the individual two-level systems, the number of two-level systems $N$ and the number of involved photons $n_{ph}$ come into play. If we would like to express the resulting hybridization in terms of an effective coupling of a single two-level system to the mode, we therefore see that we have two collective knobs to turn. The effective coupling can be increased, on the one hand, in this simplified picture by increasing the number of identical systems $N$ inside the cavity, such that we have an increase of $\sqrt{N}$. This perspective suggests that the individual matter systems (here simplified to a two-level system) are only affected by the fundamental coupling and consequently do almost not change, and the splitting is a macroscopic quantity of the resulting collective (bright) state of symmetrized superpositions. This behavior is recovered in Sec.~\ref{modelsystem}, where the unperturbed photonic system ($n_{ph}=0$) couples to many identical matter systems.\\

\noindent
On the other hand, we also see that we can increase the effective coupling by the number of involved photons. In the simplified Tavis-Cummings description, where the ground-state of the combined light-matter system is merely the tensor product of the bare electronic and photonic ground state, this is associated with higher-excited states of the combined light-matter system. However, as also seen in Sec.~\ref{modelsystem} beyond the rotating-wave approximation, even within a simplified model there is an increase of photon occupation in the ground state, which in turn can lead to an increase of the effective coupling strength. Physically this increase can be interpreted as the back-reaction of the perturbed photonic vacuum on the matter subsystem, that is, the Lamb shift. The polarization of the electromagnetic vacuum is stronger the more charges are placed within it, for example, in the single-excitation subspace the corresponding energy shift of the ground state goes as $N\lambda^2$ (see Sec.~\ref{modelsystem}). The effective increase of the coupling strength due to the vacuum polarization therefore can change the matter subsystem locally and really affect the electronic structure and the applied local few-level approximation. It is this type of increase of the effective coupling strength we consider if we scale the fundamental coupling for an individual matter system.\\

\noindent
Let us give a slightly more formal explanation of those two effects. If we consider the Green's function (propagator) of a single electron or hole $G(r_1t_1,r_2t_2)= -i \langle \mathcal{T}\hat{\Psi}(r_1t_1)\hat{\Psi}^\dagger(r_2t_2) \rangle$ (see Refs.~\cite{fetter2003, stefanucci2013} for more details), then it is influenced by the electronic environment via the Coulomb interaction (longitudinal photons) and the exchange of transversal photons. The transversal photons themselves are described by their mean-field, that is, classical, contribution and the photon-propagator  $D_\alpha(t_1,t_2)= -i \langle \mathcal{T}\Delta\hat{q}_\alpha(t_1)\Delta\hat{q}_\alpha(t_2) \rangle$, where $\Delta\hat{q}_\alpha(t_2)=\hat{q}_\alpha(t_2)-q_\alpha(t_2)$ is the deviation from the mean-field contribution. In turn, also the photon propagator is influenced by the environment, which results in a set of coupled Dyson equations  of the form
\begin{align*}
 G(1,2) &= G_0(1,2)\\&+ \int d 3 \int d 4  G_0(1,3) \Sigma(3,4) G(4,2)\\
 D(t_1,t_2) &= D_0(t_1,t_2)\\ &+ \int d 3\int d 4  D_0(t_1,t_3) \Pi(3,4) D(t_4,t_2)~,
\end{align*}
where we use the abbreviation $1\equiv (t_1,r_1)$ and correspondingly for the integrals. Furthermore $\Sigma[G,D]$ is the matter self-energy and $\Pi[G,D]$ is the polarization of the mode and $G_0$ and $D_0$ are the bare, non-interacting Green's functions of the matter and the photons, respectively. In lowest order in the matter subsystem, we can take the bare photon propagator and get a behavior that resembles the first effect of the collective interaction that goes as $\sqrt{N}$. On the other hand, the matter polarization due to the bare matter propagator will change the photonic subsystem and already in lowest order we find that the photon number scales as $\sqrt{N}$ as well, which will lead to the second collective effect. This second effect will be further influenced by the longitudinal Coulomb interaction and if we go higher in perturbation theory we will get possible non-linear enhancements as well. If we look at microscopic systems where the Coulomb interaction can be substantial, this might enhance this local effect on the matter system even further~\cite{de2018cavity}.

\section{Analytic coupling elements}\label{analytic_coupling}
\noindent
We start with the connection between the kinetic operator including the sum of all individual electronic derivatives $\sum\limits_{i=1}^{N_e} \nabla^2_{\textbf{r}_i}$ and the dipole derivative $\nabla_{R}$ 
\begin{align*}
\sum\limits_i^{N_e} \nabla^2_{\textbf{r}_i} \left[\psi^k_{\nu}(\textbf{r},\{\textbf{R}_n\}) \Phi_{k}(\textbf{q},\{\textbf{R}\})\right]~.
\end{align*}
The linear component
\begin{align*}
2\sum\limits_i^{N_e} \left[\nabla_{\textbf{r}_i}\psi^k_{\nu}(\textbf{r},\{\textbf{R}_n\})\right] \cdot \left[\nabla_{\textbf{r}_i} \Phi_{k}(\textbf{q},\{\textbf{R}\})\right]
\end{align*}
can be simplified as
\begin{align*}
\partial_{r^\mu_j} f(\textbf{R}) = \partial_{R^\mu} f(\textbf{R}) \partial_{r^\mu_j} (R^\mu_e - R^\mu_n ) = \partial_{R^\mu} f(\textbf{R}),
\end{align*}
where here $\mu=(1,2,3)$ the 3 spatial dimensions, to
\begin{align*}
2\left[\nabla_{R} \Phi_{k}(\textbf{q},\{\textbf{R}\})\right] \cdot \sum\limits_i^{N_e} \left[\nabla_{\textbf{r}_i}\psi^k_{\nu}(\textbf{r},\{\textbf{R}_n\})\right] ~.
\end{align*}
This procedure holds true for the nuclear coupling with the adjustment
\begin{align*}
\partial_{R^\mu_j} f(\textbf{R}) = \partial_{R^\mu} f(\textbf{R}) \partial_{R^\mu_j} (R^\mu_e - R^\mu_n ) = -Z_j \partial_{R^\mu} f(\textbf{R})~.
\end{align*}
The same is true for the Laplacian acting on the photonic component leading with $\sum_{i=1}^{N_e} = N_e$ to derivatives with respect to the electronic dipole.\\
The element
\begin{align*}
&\int d\textbf{q} \Phi_{l}^*(\textbf{q},\{\textbf{R}\}) \nabla_{R} \Phi_{k}(\textbf{q},\{\textbf{R}\})\\
&= \int d\textbf{q} \prod_{\alpha'}^{M_p} \phi^*_{\alpha',l}(q_{\alpha'} - q_{\alpha'}^{0}) \nabla_{R} \prod_{\alpha}^{M_p} \phi_{\alpha,k}(q_\alpha - q_\alpha^0)\\
&= \sum\limits_{\alpha}^{M_p} \int dq_\alpha \phi^*_{\alpha,l}(q_\alpha - q_\alpha^0) \nabla_{R} \phi_{\alpha,k}(q_\alpha - q_\alpha^0)
\end{align*}
can now be directly calculated with the explicit form of the HO eigenstates
\begin{align*}
&\phi_{\alpha,k}(q_\alpha - q_\alpha^0)\\ \notag &= \frac{\omega_\alpha}{\pi}^{1/4} \frac{1}{\sqrt{2^k k!}} H_k[\sqrt{\omega}(q_\alpha - q_\alpha^0)] e^{-\frac{1}{2}\omega_\alpha (q_\alpha - q_\alpha^0)^2}~,\\
&H_k[\sqrt{\omega}(q_\alpha - q_\alpha^0)] = \frac{(-1)^k}{\omega_\alpha^{\frac{k}{2}}} e^{\omega_\alpha (q_\alpha - q_\alpha^0)^2} \frac{d^k}{dq^k_\alpha} e^{-\omega_\alpha (q_\alpha - q_\alpha^0)^2}
\end{align*}
and $q_\alpha^0 = -\frac{1}{\omega_\alpha} \boldsymbol \lambda_\alpha \cdot \textbf{R}$. We could now either directly use this explicit form of the HO eigenstates or the fact, that we explicitly know the translation-operator $\hat{D}(q_\alpha^0)$ \eqref{coshifttd} such that
\begin{align*}
\nabla_{\textbf{r}_j} \phi_{\alpha,k}(q_\alpha - q_\alpha^0) &= \nabla_{\textbf{r}_j} \hat{D}_\alpha(q_\alpha^0(\{\textbf{R}\})) \phi_{\alpha,k}(q_\alpha)\\
&=  \left(\nabla_{R}\exp{\left[-i q_\alpha^0(\{\textbf{R}\}) \hat{p}_\alpha\right]}\right) \phi_{\alpha,k}(q_\alpha)\\
&= i \frac{\boldsymbol\lambda_\alpha}{\omega_\alpha} \hat{p}_\alpha \phi_{\alpha,k}(q_\alpha - q_\alpha^0)
\end{align*}
and vice versa for
\begin{align*}
\nabla_{R_j} \phi_{\alpha,k}(q_\alpha - q_\alpha^0) = -i Z_j \frac{\boldsymbol\lambda_\alpha}{\omega_\alpha} \hat{p}_\alpha \phi_{\alpha,k}(q_\alpha - q_\alpha^0)~.
\end{align*}
With $\hat{p}_\alpha = +i \sqrt{\frac{\omega_\alpha}{2}} (\hat{a}^\dagger_\alpha-\hat{a}_\alpha)$ and 
\begin{align*}
\langle l \vert \hat{a}^\dagger - \hat{a} \vert k \rangle_\alpha = \left[ \sqrt{k_\alpha+1} \delta^\alpha_{l,k+1} - \sqrt{k_\alpha} \delta^\alpha_{l,k-1} \right]~,
\end{align*}
we reach the element presented in equation \eqref{couplings2}. This involves momentum-transfer between electronic states. The same procedure for the second order term leads to \eqref{couplings3} which in contrast does not transfer momentum and represents within the same mode fluctuations and squeezings $\Delta^{lk} \sim -\hat{p}^2 \sim (\hat{a}-\hat{a}^\dagger)^2 \sim -(2\hat{a}^\dagger\hat{a}+1) + \hat{a}^\dagger\hat{a}^\dagger + \hat{a}\hat{a}$ and  couples different modes. The second oder coupling term, which corresponds to the diamagnetic $\hat{A}^2$ contribution, can be intuitively understood as renormalization of the electronic or photonic excitation. For instance the collective energetic shift $\mathcal{L}$ in Sec.~\ref{connection_qo} increases with increasing photon-number and introduces an additional detuning between excitations. This non-linearity is intuitively elucidating the discussion of App.~\ref{colcoupling}.\\

\section{Implications for periodic systems}\label{periodic}
\noindent
The initial Hamiltonian \eqref{hamiltonian} is not periodic for $\lambda > 0$ in the electronic positions $\textbf{r}$ since the photonic interaction breaks the translational invariance in this direction. 
While this problem is non-trivial to tackle directly for \eqref{hamiltonian}, it is trivial to extend the polaritonic equation \eqref{maineq} to periodic boundary conditions. The additional momentum $\nabla_{\textbf{r}}$ in the linear coupling is well suited for periodicity and the electronic surfaces $E_{\nu}(\textbf{R}_n)$ just have to be reinterpreted as polariton-Bloch quasi-particles based on the pure electronic eigenfunctions $\varphi_{\textbf{k}}(\textbf{r},\{\textbf{R}_n\}) $.  Here the momentum $\textbf{k}$ labels a continuous excitation from which the electronic bands in a periodic material emerge via back-folding into the first Brillouin-zone.\\
We present a simple example to clarify the above statement. Assume we want to solve the periodic Kohn-Sham equations for a solid within a cavity and saturated spins.
We start with a Fourier-ansatz for a single-particle orbital in the excited state $\textbf{k}$
\begin{align*}
\varphi_{\textbf{k}}(\textbf{r},\{\textbf{R}_n\}) = \sum_{\textbf{G}}^{n_{xyz}^{max}} c_\textbf{k}(\textbf{G}) e^{i(\textbf{k}+\textbf{G})\cdot \textbf{r}}
\end{align*}
where $\textbf{G}$ is the inverse lattice vector parametrically depending on the nuclear configuration and $n_{xyz}^{max}$ the corresponding cut-off. By construction of Kohn-Sham theory \cite{hohenberg1964,kohn1965}, the single particle orbitals are solved in a non-linear self-consistent procedure using an effective Kohn-Sham potential that mimics the many-body electronic interactions. The set of $\varphi_{\textbf{k}},~\overline{\varepsilon}_{\textbf{k}} (\{\textbf{R}_n\}),~c_\textbf{k}(\textbf{G}) $ is then given as solution to
\begin{align*}
&\overline{\varepsilon}_{\textbf{k}} (\{\textbf{R}_n\}) c_\textbf{k}(\textbf{G}) = \sum_{\textbf{G}'}^{n_{xyz}^{max}}\bigg( \left[  \frac{\vert \textbf{k} + \textbf{G}' \vert^2}{2}  + V_{nn}(\{\textbf{R}_n\}) \right] \delta_{\textbf{G} \textbf{G}'}  \\
& +    v_{KS}(\textbf{G}-\textbf{G}') \bigg) c_\textbf{k}(\textbf{G}')~.
\end{align*}
This defines a continuous basis with eigenvalues related to momenta $\textbf{k}$ which are now used as input in equation \eqref{maineq} or \eqref{rothaan_hall}. As the simplest example, the single photon polariton is given as
\begin{widetext}
	\begin{align*}
	\left(
	\begin{matrix}
	\overline{\varepsilon}_{\textbf{k}} + \varepsilon_0 - \frac{N_e}{2} \Delta^{00} & 0 & 0 & \nabla^{\textbf{k}\textbf{k}'} \cdot \nabla^{10} \\ 
	0 & \overline{\varepsilon}_{\textbf{k}} + \varepsilon_1 - \frac{N_e}{2} \Delta^{11} & \nabla^{\textbf{k}\textbf{k}'} \cdot \nabla^{01} & 0 \\
	0 & \nabla^{\textbf{k}'\textbf{k}} \cdot \nabla^{10} & \overline{\varepsilon}_{\textbf{k}'} + \varepsilon_0 - \frac{N_e}{2} \Delta^{00} & 0 \\
	\nabla^{\textbf{k}'\textbf{k}} \cdot \nabla^{01} & 0 & 0 & \overline{\varepsilon}_{\textbf{k}'} + \varepsilon_1 - \frac{N_e}{2} \Delta^{11}
	\end{matrix}
	\right)
	\end{align*}
\end{widetext}
where the momenta are connected according to
\begin{align*}
\nabla^{\textbf{k}\textbf{k}'} &= \sum_{\textbf{G},\textbf{G}'} c^*_\textbf{k}(\textbf{G}) c_{\textbf{k}'}(\textbf{G}') i(\textbf{k}' + \textbf{G}') \delta_{\textbf{k} + \textbf{G}, \textbf{k}' + \textbf{G}'}~.
\end{align*}
Within the long-wavelength approximation, that is, there is no momentum-transfer between photon and electron, the electronic momentum is a conserved quantity up to Umklapp scattering. Note that $\nabla^{\textbf{k}\textbf{k}} = 0$ for real functions. As a consequence, the quantized photonic nature splits the bare electronic excited bands into polaritonic bands, that is, similar to the interpretation of upper and lower polariton as for a two-level system in Sec.~\ref{modelsystem}.\\
Solving this equation will already imprint first quantum features in periodic materials from a natural and simple perspective.
We want to emphasize here, that this does not describe the correct many-body excitations since each of the Kohn-Sham orbitals is coupled to the field instead of many-body states as the Kohn-Sham Slater-determinants. The above procedure is consequently the one of an effective single particle one. Calculations for realistic two-dimensional systems are under investigation~\cite{ronca2018twodimensional}. The correct description would involve coupling of many-body eigenstates corresponding to poles of the spectral function~\cite{ullrich2011}.

\section{Time-dependent Born-Huang expansion}\label{timedependent}
\noindent
For a wavefunction with trivial equilibrium time-dependence
\begin{align*}
\Psi_{i}(\textbf{R}_n, \textbf{r}, \textbf{q},t) = e^{-iE_i t}\Psi_{i}(\textbf{R}_n, \textbf{r}, \textbf{q})~,
\end{align*}
that is, a time-independent Hamiltonian, the time-argument can be purely absorbed by the nuclear component. Therefore, the only change appearing is the substitution in \eqref{nsub} of $E_i \rightarrow i\partial_t$ and $\chi^{\mu}_{i}(\textbf{R}_n)\rightarrow\chi^{\mu}_{i}(\textbf{R}_n,t)$. The nuclear wavepacket is then moving on frozen polaritonic surfaces. One possible realization is an instantaneous Frank-Condon transition which does not change the eigenstates $\Psi_{i}$ but results in a superposition of different eigenstates as initial configuration $\Psi(t) = \sum_i c_i(t) \Psi_i(t=0)$ \cite{ullrich2011}.\\[1\baselineskip]
Non-equilibrium dynamics, where the Hamiltonian is time-dependent, is of vast interest in chemistry and physics. Prototypical examples are the coupling to an external Laser-field $ \hat{\textbf{R}}\cdot \textbf{E}(t) $ in the nuclear-electronic component 
or the displacements of the photonic variable $ \hat{q}_\alpha \rightarrow \hat{q}_\alpha - j^{\alpha}_{ext}(t) $ by an external current. Similar to the exact equilibrium reformulation \eqref{fullseparated1}-\eqref{fullseparated4}, we can derive the non-equilibrium analogue with the Born-Huang ansatz
\begin{align*}
&\Psi_{i}(\textbf{R}_n, \textbf{r}\boldsymbol, \textbf{q},t)\\ &= \sum_{\mu, k = 0}^{\infty} \chi^{\mu}_{i}(\textbf{R}_n,t) \psi^k_{\mu}(\textbf{r}, \{\textbf{R}_n\},t) \Phi_{k}(\textbf{q},\{\textbf{R}\},t)~.
\end{align*}
While the decomposition of the Hamiltonian is not affected by this, the partial time-derivative results in three components
\begin{align}\label{tdbh}
&\sum_{l=0}^{\infty} \left(\langle \psi^{l}_{\nu}(\{\textbf{R}_n\},t)\vert \langle\Phi_{l}(\{\textbf{R}\},t)\vert\right) i\partial_t \vert \Psi_{i}(\textbf{R}_n,t)\rangle\notag\\
&= i\partial_t \chi^{\nu}_{i}(\textbf{R}_n,t)\\
&+ \sum_{\mu,l= 0}^{\infty} \chi^{\mu}_{i}(\textbf{R}_n,t) \langle\psi^l_{\nu}(\{\textbf{R}_n\},t)\vert i\partial_t \notag \vert\psi^l_{\mu}(\{\textbf{R}_n\},t)\rangle_e\notag\\
&+ \sum_{\mu,l, k = 0}^{\infty} \chi^{\mu}_{i}(\textbf{R}_n,t) \langle\psi^l_{\nu}(\{\textbf{R}_n\},t)\vert\psi^k_{\mu}(\{\textbf{R}_n\},t)\rangle_e\notag\\
& \qquad\qquad\times \langle \Phi_{l}(\{\textbf{R}\},t) \vert i\partial_t \vert \Phi_{k}(\{\textbf{R}\},t)\rangle_p \notag~.
\end{align}
There are different possibilities how the time dependence can be re-casted efficiently. One possible approach involves the definition of a parametric phase $e^{i\overline{\varepsilon}_k(\{\textbf{R}\},t)}\Phi_k(\textbf{q},\{\textbf{R}\},t)$ \cite{cederbaum2008born}. Since each subsystem is invariant under a global phase which in this case depends parametrically on the trajectories, the combined full wavefunction can be constructed with those parametric phases. Notice that this gives rise to additional non-adiabatic coupling elements $\nabla \overline{\varepsilon}_k(\{\textbf{R}\},t)$.\\
Each time-derivative of \eqref{tdbh} can be assigned to one Hamiltonian equation of each species, for example
\begin{align} \label{phsubtd}
\begin{split}
&\hat{H}_{BO}^{ph}(\{\textbf{R}(t)\},t) \Phi_{k}(\textbf{q},\{\textbf{R}\},t) = i\partial_t \Phi_{k}(\textbf{q},\{\textbf{R}\},t)~.
\end{split}
\end{align}
The parametric dependence in the Hamiltonian and wavefunctions is now time-dependent, that is, the total dipole in the photonic equation \eqref{phsubtd} for example involves parametric classical trajectories $\textbf{R}(t)$ such that equation \eqref{phsubtd} could be intuitively solved not just for a set of values $\textbf{R}$ as before but for a set of trajectories $\textbf{R}(t)$. Additionally, the non-adiabatic coupling elements change over time.
Within the long-wavelength assumption, the photonic system can be solved analytically (see Sec.~\ref{photonsub}).

\bibliography{01_light_matter_coupling} 

\end{document}